\documentclass[fleqn,usenatbib]{mnras}
\usepackage{times}
\usepackage{amssymb}
\usepackage{amsmath}
\usepackage{pifont}
\usepackage{graphics}
\usepackage{xcolor}
\usepackage{epsfig}
\usepackage{xspace,comment}
\usepackage[T1]{fontenc}
\usepackage{ae,aecompl}
\newcommand{\thisstar}{KELT J072709+072007}

\newcommand{\msun}{M\textsubscript{$\odot$}}    
\newcommand{\rsun}{R\textsubscript{$\odot$}}    
\newcommand{\lsun}{L\textsubscript{$\odot$}}    
\newcommand{\MJ}{M\textsubscript{J}}          
\newcommand{\RJ}{R\textsubscript{J}}          

\newcommand{\feh}{\ensuremath{\left[{\rm Fe}/{\rm H}\right]}\xspace}
\newcommand{\teff}{\ensuremath{T_{\rm eff}}\xspace}
\newcommand{\logg}{\ensuremath{\log g_*}\xspace}
\newcommand{\vsini}{\ensuremath{v \sin I_*}\xspace}

\newcommand{\MS}{\ensuremath{\,\msun}\xspace}
\newcommand{\RS}{\ensuremath{\,\rsun}\xspace}
\newcommand{\LS}{\ensuremath{\,\lsun}\xspace}

\newcommand{\kms}{\,km\,s$^{-1}$\xspace}

\title[{\bf Characterization of a Rare B-M Eclipsing Binary}]{An Extreme-mass Ratio, Short-period Eclipsing Binary Consisting of a B Dwarf Primary and a Pre-main Sequence M Star Companion Discovered by KELT}

\author[D. J. Stevens et al.]{ Daniel J. Stevens,$^{1,2,\dagger}$\thanks{E-mail: danstevens@psu.edu} George Zhou,$^{3,\star\star}$ Marshall C. Johnson,$^{4}$ Aaron C. Rizzuto,$^{5}$ \newauthor Joseph E. Rodriguez,$^{3}$ Allyson Bieryla,$^{3}$ Karen A. Collins,${^3}$ Steven Villanueva, Jr.,$^{6}$ \newauthor Jason T. Wright,$^{1,2}$ B. Scott Gaudi,$^{7}$ David W. Latham,$^{3}$ Thomas G. Beatty,$^{8}$ \newauthor Michael B. Lund,$^{9}$ Robert J. Siverd,$^{10}$ Adam L. Kraus,$^{5}$ Patcharapol Wachiraphan,$^{11}$ \newauthor  Perry Berlind,$^{3}$ Michael L. Calkins,$^{3}$ Gilbert A. Esquerdo,$^{3}$ John F. Kielkopf,$^{12}$ \newauthor  Rudolf B. Kuhn,$^{13,14}$ Mark Manner,$^{15}$ Joshua Pepper,$^{16}$ and Keivan G. Stassun$^{17,18}$
\\
$^{1}$Center for Exoplanets and Habitable Worlds, The Pennsylvania State University, 525 Davey Lab, University Park, PA 16802, USA\\
$^{2}$Department of Astronomy \& Astrophysics, The Pennsylvania State University, 525 Davey Lab, University Park, PA 16802, USA\\
$^{3}$Center for Astrophysics | Harvard \& Smithsonian, 60 Garden St, Cambridge, MA 02138, USA\\
$^{4}$Las Cumbres Observatory, 6740 Cortona Drive, Suite 102, Goleta, CA 93117, USA\\
$^{5}$Department of Astronomy, The University of Texas at Austin, 2515 Speedway, Stop C1400, Austin, TX 78712, USA\\
$^{6}$Department of Physics and Kavli Institute for Astrophysics and Space  Research, Massachusetts Institute of Technology,\\ \ Cambridge, MA 02139, USA\\
$^{7}$Department of Astronomy, The Ohio State University, 140 W. 18th Ave, Columbus, OH 43210, USA\\
$^{8}$Department of Astronomy and Seward Observatory, University of Arizona, Tucson, AZ 85721, USA\\
$^{9}$Caltech IPAC -- NASA Exoplanet Science Institute 1200 E. California Ave, Pasadena, CA 91125, USA\\
$^{10}$Gemini Observatory, Northern Operations Center, 670 N. A'ohoku Place, Hilo, HI 96720, USA\\
$^{11}$Department of Physics, Faculty of Science, Mahidol University, Bangkok 10400, Thailand\\
$^{12}$Department of Physics and Astronomy, University of Louisville, Louisville, KY 40292, USA\\
$^{13}$South African Astronomical Observatory, P.O. Box 9, Observatory 7935, South Africa\\
$^{14}$Southern African Large Telescope, PO Box 9, Observatory, 7935, Cape Town, South Africa\\
$^{15}$Spot Observatory, Nashville TN, 37206, USA\\
$^{16}$Department of Physics, Lehigh University, 16 Memorial Drive East, Bethlehem, PA 18015, USA \\
$^{17}$Vanderbilt University, Department of Physics \& Astronomy, 6301 Stevenson Center Ln., Nashville, TN 37235, USA\\
$^{18}$Fisk University, Department of Physics, 1000 18th Ave. N., Nashville, TN 37208, USA\\
$^{\dagger}$Eberly Research Fellow\\
$^{\star\star}$Hubble Fellow}
\date{ Accepted 2020 October 1. Received 2020 September 28; in original form 2019 October 18.} 
\pubyear{2020} 
\begin{document}

\label{firstpage}
\pagerange{\pageref{firstpage}--\pageref{lastpage}}
\maketitle
\begin{abstract}

We present the discovery of \thisstar\ (HD 58730), a very low mass ratio ($q \equiv M_2/M_1 \approx 0.07$) eclipsing binary (EB) identified by the Kilodegree Extremely Little Telescope (KELT) survey. We present the discovery light curve and perform a global analysis of four high-precision ground-based light curves, the Transiting Exoplanets Survey Satellite (TESS) light curve, radial velocity (RV) measurements, Doppler Tomography (DT) measurements, and the broad-band spectral energy distribution (SED). Results from the global analysis are consistent with a fully convective ($M_2 = 0.22 \pm 0.02\ \MS)$ M star transiting a late-B primary ($M_1 = 3.34^{+0.07}_{-0.09}\ \MS;\ T_{\rm eff,1} = 11960^{+430}_{-520}\ {\rm K}$). We infer that the primary star is $183_{-30}^{+33}$ Myr old and that the companion star's radius is inflated by $26 \pm 8\%$ relative to the predicted value from a low-mass isochrone of similar age. We separately and analytically fit for the variability in the out-of-eclipse TESS phase curve, finding good agreement between the resulting stellar parameters and those from the global fit. Such systems are valuable for testing theories of binary star formation and understanding how the environment of a star in a close-but-detached binary affects its physical properties. In particular, we examine how a star's properties in such a binary might differ from the properties it would have in isolation. \end{abstract}
\begin{keywords}
stars: early-type -- stars: low-mass -- stars: pre-main-sequence -- (stars:) binaries: eclipsing -- eclipses
\end{keywords}

\section{Introduction}

Surveying binary stars and characterizing their masses, radii, orbital periods/separations, and orbital eccentricities can tell us about how intermediate-mass stars and their lower-mass companions form and evolve, as well as whether different formation mechanisms dominate different regions of parameter space. Recent studies of high-mass ($\gtrsim 10\MS$) star formation have demonstrated that fragmentation of a massive protostar's disk can lead to the formation of low-mass stars and brown dwarf companions (hereafter extreme mass ratio binaries, or EMRBs) with 10-100 au orbital separations. Episodic accretion onto the protostar, with long intra-burst periods, catalyzes the formation of a low-mass stellar companion \citep{Stamatellos:2011}. Episodic accretion onto young protostars appears to be a common occurrence as observed in the FU Orionis class of FGK stars (e.g. \citealt{Herbig:1977,Dopita:1978, Reipurth:1989,Greene:2008,Peneva2010}; see also \citealt{Hartmann1996} for a review). However, these efforts have neither focused on more massive ($3-10\MS$) primary stars nor explored stars in binaries with orbital separations below $\sim 10$ au. 

There have been recent efforts to discover and characterize EMRBs. \citet{Gullikson2013} developed the direct spectral detection method, which is sensitive to binaries with mass ratios above $q \approx 0.1$ and small projected separations due to the lack of an inner working angle. They applied it to identify G- and K-type companions to early B-type stars. \citet{Gullikson:2016} applied this technique to 341 B and A stars, estimating a survey completeness of $20-30\%$ at a mass ratio $q \sim 0.05$. \citet{Evans:2011} found potential low-mass (here, $M \lesssim 1.4 \MS$) companions to late B-type stars by looking for coincident X-ray detections, arguing that late B-type stars are typically X-ray quiet (excluding magnetic chemically peculiar stars) and so any detected X-ray flux would be emitted by an active, lower-mass companion.

Similarly, the Volume-limited A-Star (VAST) survey \citep{DeRosa:2011} identified EMRB candidates via AO imaging followup of intermediate-mass stars with coincident X-ray detections in ROSAT data \citep{Voges:1999}. VAST also identified EMRBs and EMRB candidates via common proper motion and AO orbital analyses \citep{DeRosa:2014}. The typical projected separation scale of the AO-detected binaries is $10-10^4$ au.

\citet{Moe2015} found 18 young ($\lesssim 8$ Myr) eclipsing EMRBs consisting of early-B type primaries and G/K-type companions on short ($P\approx 3-18$ d) orbits in the Large Magellanic Cloud. Using this sample, they infer that $2\pm 6 \% $ of B-type stars have short-period companions with $0.06 \lesssim q \lesssim 0.25$\ but their sample does not include any B-M binaries.

Finding very short-period ($P \lesssim 10$ days), very-low-mass ($M_2 \lesssim 0.3 \MS$) companions to main-sequence intermediate-mass stars proves difficult for a number of reasons. In such binaries, the M dwarf contributes negligibly to the total flux: in a B8V-M3V binary with $q \approx 0.1$, the M dwarf contributes $0.1\%$ of the bolometric flux; the contribution is lower in optical wavelengths. As such, the aforementioned methods to survey intermediate-mass stars for dim companions cannot probe both small separations ($\lesssim 10$ au) and very low flux ratios (corresponding to low mass ratios).

In particular, spectroscopic detections of such binaries are hindered by a few additional complications. Intermediate-mass main-sequence stars are often rapidly rotating. These stars lie above the Kraft break, roughly $\teff \sim 6250$ K \citep{Kraft:1967}, where they no longer have thick convective envelopes, leading to weaker magnetic braking of the stellar rotation. This lack of spin-down leads to rapid rotation of the star throughout its lifetime. The high rotation rates significantly broaden the primary star's few strong spectral lines; a low-mass stellar companion can still be identified by high-amplitude, periodic variations in the primary star's radial velocity (RV), but the rotational broadening and paucity of absorption lines diminishes the achievable precision.

Exoplanet transit surveys offer one promising way of identifying low mass-ratio eclipsing binaries (EBs). Since the radius of a late-M dwarf is approximately the same as the radius of Jupiter, and since the transit depth is proportional to the companion-to-primary radius ratio, both planetary and late-type stellar companions would produce comparable transit/eclipse signals around an intermediate-mass host star. RV follow-up would then be required to identify the system as a stellar binary. The Eclipsing Binaries with Low Mass (EBLM) project out of the Wide-Angle Search for Planets (WASP) survey (\citealt{Triaud2013,Gomez2014,Triaud2017}) has characterized over 100 single-lined EBs, including a handful of EMRBs \citep{vonBoetticher2019} such as EBLM J0555-57 \citep{vonBoetticher2017}. However, only a couple of these EBLMs have primary stars whose effective temperatures are consistent with the A spectral type, and none that fall in the B spectral type effective temperature range, likely due in part to the aforementioned observational difficulties. Thus, new discoveries of EMRBs with intermediate-mass primary stars can add substantially to our knowledge of these systems.

Although the previously mentioned challenge of measuring precise RVs for rapidly rotating stars becomes even more difficult when trying to measure the smaller RV signal induced by a planetary-mass companion, there have been an increasing number of confirmations of planets transiting very rapidly rotating stars through the use of Doppler Tomography (DT), in which the transit-induced perturbation to the rotationally broadened spectral line profile is resolved spectroscopically (e.g. \citealt{Collier:2010b,Bieryla:2015}). We list the planets around rapid rotators (\vsini $>$ 70 km/s) with DT, or spectroscopic transit, observations in Table \ref{tab:dt}. This recent focus on obtaining detailed spectroscopic follow-up observations of transit-identified planet candidates around hot stars demonstrates the planet-hunting community's sensitivity to low-mass stellar companions around hot, intermediate-mass stars and the promise of using existing transit surveys and their follow-up infrastructures to find and characterize them.

\begin{table*}
\centering
\caption{Confirmed Planets around Rapid Rotators with Doppler Tomography Observations}
\label{tab:dt}
\begin{tabular}{lcl}
  \hline
  \hline
Name & \vsini\ (km/s) & Reference\\ 
\hline
HAT-P-69b & 77 & \citet{Zhou:2019}\\
Kepler-13Ab & 77 & \citet{Szabo:2011,Johnson:2014}\\
KELT-19b & 85 & \citet{Siverd2018}\\
WASP-33b & 86 & \citet{Collier:2010b}\\
HAT-P-70b & 100 & \citet{Zhou:2019}\\
WASP-189b & 100 & \citet{Anderson:2018}\\
HAT-P-57b & 102 & \citet{Hartman:2015}\\
MASCARA-1b & 109 & \citet{Talens:2017}\\ 
KELT-9b& 111 & \citet{Gaudi:2017}\\ 
KELT-25b & 114 & \citet{RodriguezMartinez:2020}\\
KELT-20b/MASCARA-2b & 116& \citet{Lund:2017,Talens:2018}\\
KELT-26b & 123 & \citet{RodriguezMartinez:2020}\\
KELT-21b & 146& \citet{Johnson:2018}\\
\hline
\hline
\end{tabular}
\end{table*}

In this paper, we present the discovery and characterization of \thisstar, an EB consisting of a late-B type primary and an M-type companion with a short, 3.6-day orbital period. We describe the discovery along with our light curve, RV, DT, and adaptive optics (AO) observations in Section \ref{sec:disc}. We present the physical characterization of the system using these observations, broad-band flux measurements, and stellar models in Section \ref{sec:GlobalFit}. We discuss \thisstar's significance in the context of intermediate mass-low mass stellar binaries in Section \ref{sec:discussion}.

\section{Discovery and Follow-Up Observations}
\label{sec:disc}
\subsection{KELT-South and KELT-North}
\thisstar\ ($\alpha = 07^h27^m09^s40$, $\delta = +07\degr 20\arcmin07\farcs40$; J2000) lies in the KELT field KJ06 ($\alpha = 7^h40^m12^s$, $\delta = +3\degr$; J2000), which was observed jointly by both KELT-North and KELT-South \citep{Pepper:2007,Pepper:2012,Pepper:2018}. The KELT survey telescopes observed KJ06 a total of 5184 times between 2010 March and 2015 May, with 2024 observations from KELT-North and 3160 from KELT-South. After reducing the KELT-North and KELT-South observations using the standard KELT data reduction routines described in \citet{Siverd:2012} and \citet{Kuhn:2016}, respectively, and the planet candidate selection routines described in \citet{Collins2018}, we identified \thisstar\ as a transiting planet candidate. We also identified \thisstar\ as part of a focused search for planets around hot stars, processing the light curves as described in \citet{Zhou:2016k17}. We found a period of about 3.6 d, a 5.6 hr transit duration, and a 7 mmag primary transit depth for the planet candidate.

\begin{figure}
    \centering
    \includegraphics[width=1\linewidth]{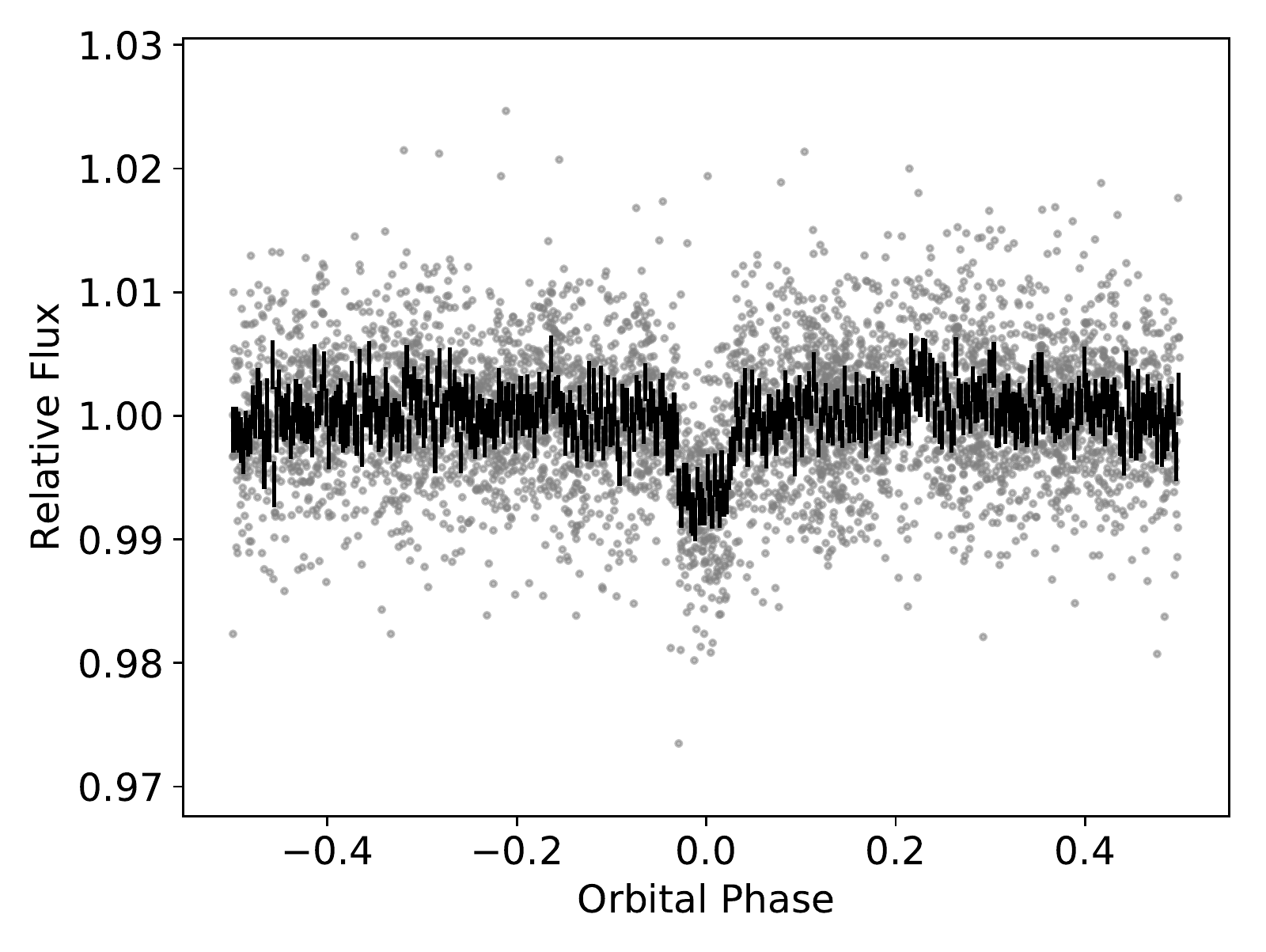}
    \caption{The discovery light curve from the KELT survey. The grey points show the discovery light curve; the black points show the same data, binned at intervals of 0.0025 in phase. The KELT discovery dataset was not used to constrain the planet parameters in the global fit (Section~\ref{sec:GlobalFit}). The transit depth from the discovery light curves are diluted due to the application of TFA, which acts to dampen any modulation in the light curve.}
    \label{fig:KS_discoverylc}
\end{figure}

The combined KELT-North and KELT-South discovery light curve, phased-folded on the best-fit ephemeris, is shown in Figure~\ref{fig:KS_discoverylc}. Due to KELT's pixel scale -- $23$ arcseconds per pixel -- the KELT light curve transits are susceptible to dilution from neighboring stars that fall within the aperture used. 
The broadband magnitudes and other stellar properties are listed in Table \ref{tab:hostprops}.

\begin{table*}
\centering
\scriptsize
\caption{Magnitudes and Kinematics of \thisstar}
\label{tab:hostprops}
\begin{tabular}{llccc}
  \hline
  \hline
  Parameter & Description &Value & Source & Reference(s) \\
  \hline
Names & & \thisstar & & \\
&   &  TIC 425223388 &  & \\
& 	&  TYC 177-95-1	& 		&	\\
& 	& 2MASS J07270942+0720074 	& 	&\\
&	& HD 58730	&	&\\
			&					&				&		&			\\
$\alpha_{J2000}$	&Right Ascension (RA)& 07:27:09.40		& Tycho-2	& \citet{Hog:2000}	\\
$\delta_{J2000}$	&Declination (Dec)& +07:20:07.40			& Tycho-2	& \citet{Hog:2000}	\\
			&					&				&		&			\\
$B$			&Johnson $B$-band magnitude & 8.778 $\pm$ 0.008		& APASS	& \citet{Henden:2015}	\\
$V$ &  Johnson $V$-band magnitude & 8.879 $\pm$ 0.009 & APASS & \citet{Henden:2015} \\
$g'$ &Sloan $g'$-band magnitude & $8.754 \pm 0.055$ & APASS & \citet{Henden:2015}\\
$r'$ & Sloan $r'$-band magnitude   & $9.011 \pm 0.014$ 		&   APASS	&   \citet{Henden:2015}	\\
$i'$ &Sloan $i'$-band magnitude & 9.315 $\pm 0.018$ & APASS & \citet{Henden:2015} \\
$G$ & {\it Gaia} $G$-band magnitude & $8.836 \pm 0.001$ & {\it Gaia} DR2 & \citet{Gaia:2018} \\
$G_{BP}$ & {\it Gaia} $G_{BP}$-band magnitude & $8.806 \pm 0.002$ & {\it Gaia} DR2 & \citet{Gaia:2018} \\
$G_{RP}$ & {\it Gaia} $G_{RP}$-band magnitude & $8.904 \pm 0.001$ & {\it Gaia} DR2 & \citet{Gaia:2018} \\

J			&2MASS magnitude& 9.019 $\pm$ 0.021		& 2MASS 	& \citet{Cutri:2003, Skrutskie:2006}	\\
H			&2MASS magnitude& 9.056 $\pm$ 0.023	& 2MASS 	& \citet{Cutri:2003, Skrutskie:2006}	\\
K			&2MASS magnitude& 9.054 $\pm$ 0.021		& 2MASS 	& \citet{Cutri:2003, Skrutskie:2006}	\\
			&					&				&		&			\\
$\pi_p$ & Parallax$^\dagger$ (mas) & $1.6250 \pm 0.0785$ & {\it Gaia} DR2 & \citet{Gaia:2018,Lindegren:2018}\\
$\mu_{\alpha}$		& Proper Motion in RA (mas yr$^{-1}$)	& -3.016
 $\pm$ 0.138	& {\it Gaia} DR2		& \citet{Gaia:2018, Lindegren:2018} \\
$\mu_{\delta}$		& Proper Motion in Dec (mas yr$^{-1}$)	&  -0.989 $\pm$ 0.121	& {\it Gaia} DR2		& \citet{Gaia:2018,Lindegren:2018} \\
RV & Absolute Radial Velocity (\kms) & 11.15 $\pm 0.13$ & &  This work \\
Distance & Distance (pc) & $615\pm30$ &  &  This work \\
U$^*$ & Space motion (\kms) &   -3.14 $\pm$ 0.25 & &  This work \\
V & Space motion (\kms) & 8.08 $\pm$ 0.32 & & This work \\
W & Space motion (\kms) & -0.32 $\pm$ 0.58 &  &  This work \\	
 \hline
\hline
\end{tabular}
\begin{flushleft}
\begin{center}
 \footnotesize \textbf{\textsc{NOTES}}\\
\footnotesize $^{*}$U is positive in the direction of the Galactic Center.\\
\footnotesize $^{\dagger}$Corrected for the $0.0820 \pm 0.033$~mas systematic offset found by \cite{Stassun:2018}.
\end{center}
\end{flushleft}
\end{table*}

\subsection{Photometric Follow-up}
\label{sec:Follow-up_Photometry}
\begin{figure}
\includegraphics[width=1\linewidth]{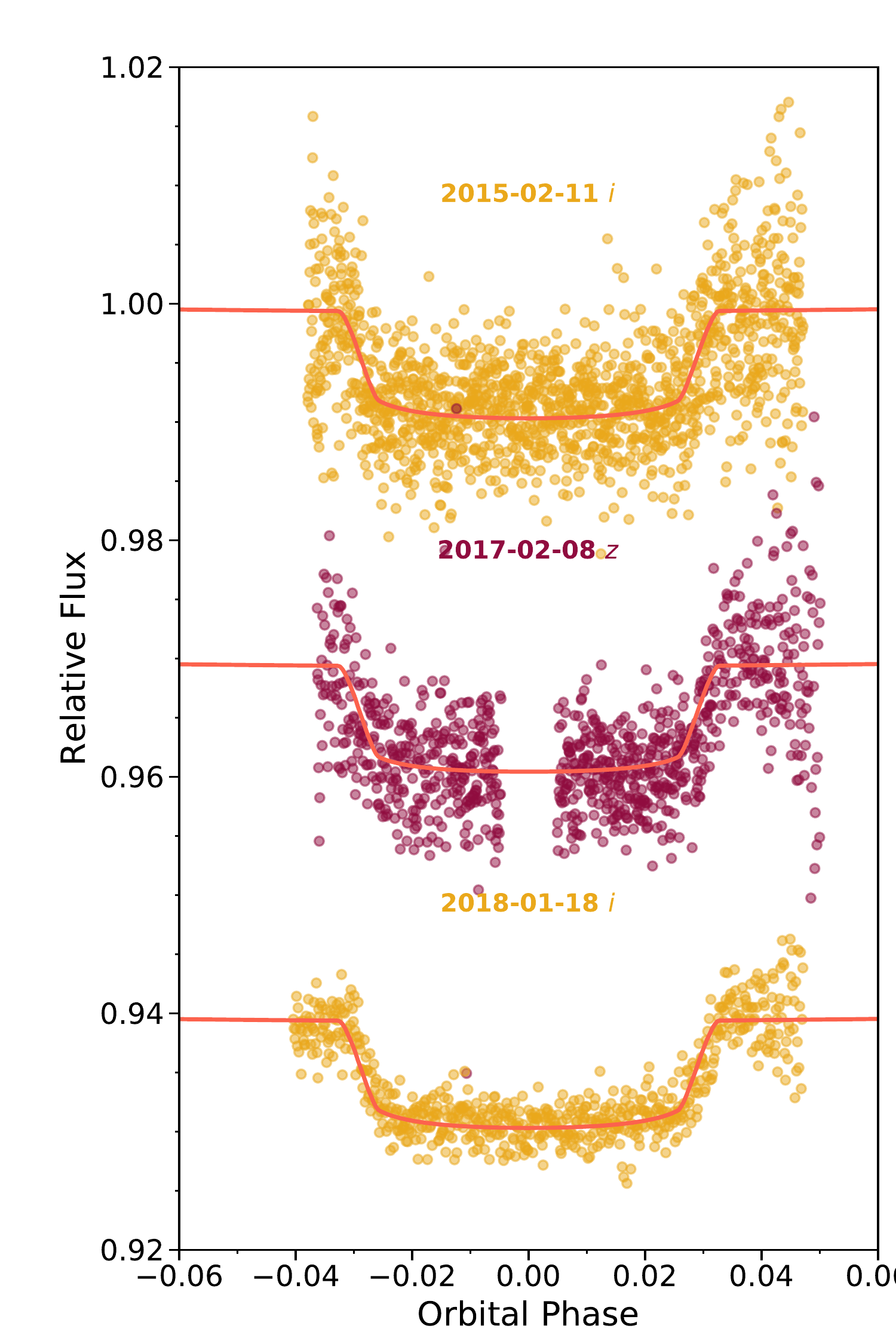}
\caption{The individual KELT follow-up Network observations of \thisstar. The observations are from KeplerCam (top two light curves) and ULMT (bottom light curve).}
\label{fig:All_light curve} 
\end{figure}

Our analysis includes three higher precision and higher spatial resolution photometric follow-up observations of \thisstar\ from the KELT Follow-up Network (KELT-FUN; \citealt{Collins2018}). These datasets are uniformly reduced using AstroImageJ \citep{Collins:2016}. We present these light curves in Figure~\ref{fig:All_light curve}. A description of each observatory is below. 

We observed one full transit in the $i'$ band and one partial transit in the $z'$ band with KeplerCam on the 1.2m telescope at the Fred Lawrence Whipple Observatory (FLWO) on UT 2015 Feb 11 and 2017 Feb 08, respectively. KeplerCam has a single $4{\rm K} \times 4{\rm K}$ Fairchild CCD with $0\farcs366$ pixel$^{-1}$ and a field of view of $23\arcmin.1 \times 23\arcmin.1$.

We observed a full transit in the $i'$ band with the 0.6m University of Louisville Manner Telescope (ULMT) on UT 2018 Jan 18. ULMT has a $4{\rm K} \times 4{\rm K}$ SBIG STX-16803 CCD camera with with a $26\arcmin.8 \times 26\arcmin.8$ field of view and a pixel scale of $0\farcs39$ pixel$^{-1}$.

\subsection{Space-based light curve from \emph{TESS}}
\label{sec:tess}

\thisstar\ (TIC 425223388) was observed by the Transiting Exoplanet Survey Satellite (TESS; \citealt{Ricker:2014}) via the 30-minute cadence Full Frame Images (FFI). \emph{TESS} observed this target in Sector 7, which ran from 2019 Jan 8 to 2019 Feb 1. Light curves were extracted from the calibrated FFIs produced by the \emph{TESS} Science Processing Operations Center \citep{2016SPIE.9913E..3EJ}, and downloaded from the MAST archive via the \emph{lightkurve} package \citep{2019AAS...23310908B}. Apertures were drawn around the target star encompassing pixels with fluxes brighter than 68\% of the surrounding background pixels. 

\begin{figure*}
    \centering
    \includegraphics[width=1.\textwidth,trim=10 0 0 10,clip]{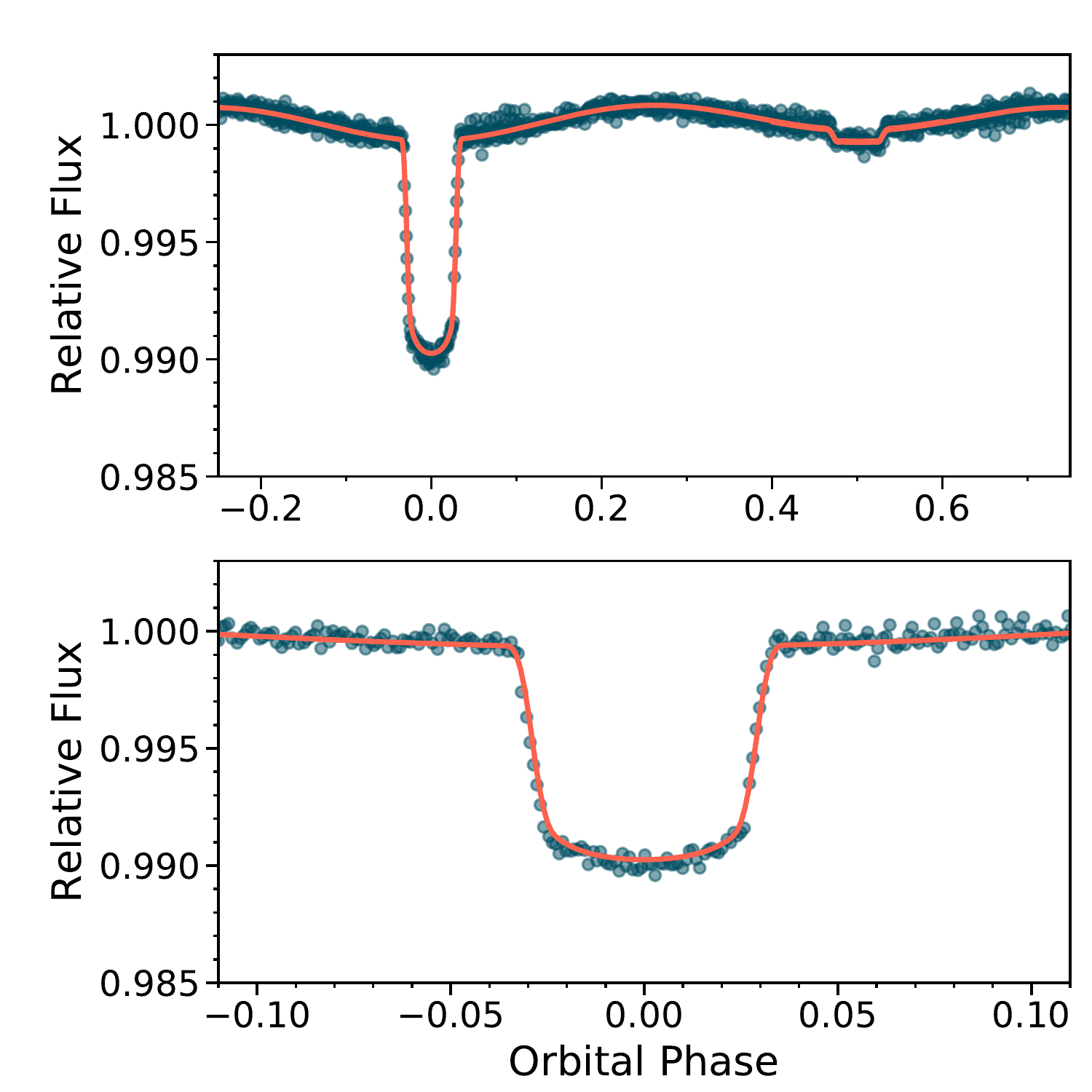}
    \caption{TESS light curve of \thisstar\ (grey points) and best-fit model (red line) phase-folded to the best-fit ephemeris from the global fit in Section \ref{sec:GlobalFit}. The 1\% transit and sub-mmag secondary eclipse are shown at phases 0.0 and 0.5, respectively, and out-of-eclipse (i.e. neither during primary nor secondary eclipse)} variability is visible, reaching maxima at orbital quadrature.
    \label{fig:tess}
\end{figure*}

The detrended and phase-folded \emph{TESS} light curve is shown in Figure~\ref{fig:tess}. The light curve reveals flux variations in-phase with the orbital period, as well as a $\sim 1\,$mmag secondary eclipse centered at phase 0.5. The detrended light curve is used in the global modeling of the system described in Section~\ref{sec:GlobalFit}. 

\subsection{Spectroscopic Follow-up}
\label{sec:spec_fu}

A series of spectroscopic follow-up observations were performed to characterize the atmospheric properties and RV variations of \thisstar. The observations are described in more detail below. The RVs used in the RV-orbit fit in Section \ref{sec:GlobalFit} are presented in Table \ref{tab:RVs}.

\begin{table*}
{
    \caption{Relative radial velocities for \thisstar}
    \label{tab:RVs}
    \centering
    \begin{tabular}{rrrrl}
    \hline\hline
BJD (UTC) & Relative RV (km\,s$^{-1}$) & RV Error (km\,s$^{-1}$) & Facility\\ 
\hline
2457685.9113403000 & -0.6009060279 & 4.0677105264 & McDonald\\
2457687.8970350400 & 8.7425612267 & 7.24292216548 & McDonald\\
2457735.8283281601 & -9.5724721306 & 1.7705572594 & McDonald\\
2457736.0179607598 & -13.9713671182 & 2.8282088599 & McDonald\\
2457736.8343254901 & 4.8502566195 & 1.3959596723 & McDonald\\
2457737.0238991501 & 5.3144305824 & 6.7067976622 & McDonald\\
2457737.7891493500 & 10.2310391586 & 1.7519105563 & McDonald\\
2457738.0249463399 & 7.1177491999 & 5.1887427262 & McDonald\\
2458141.6461992501 & -10.2571338828 & 2.6806482424 & McDonald\\
2458141.7929965700 & -1.3932320422 & 1.1184171216 & McDonald\\
2458141.9253813098 & -1.5366440476 & 4.0366142280 & McDonald\\
2458142.6460697600 & 10.6568527114 & 2.4456213398 & McDonald\\
2458142.8071196298 & 8.0439209804 & 5.8027919365 & McDonald\\
2458143.5954316799 & 16.9617061663 & 2.6376446047 & McDonald\\
2458143.8773542200 & 11.2698734912 & 3.0631777970 & McDonald\\
2458135.9256989998 & 25.9221434417 & 3.4605657655 & TRES\\
2458137.7783320001 & -0.1368383604 & 0.6053444298 & TRES\\
2458138.7431529998 & 12.8631399347 & 2.7720804962 & TRES\\
2458139.9436180000 & 21.8087696378 & 4.0975249307 & TRES\\
2458140.8006879999 & -0.8210044787 & 1.0148077119 & TRES\\
2458141.7888600002 & -2.4533205171 & 0.4044581804 & TRES\\
2458142.7684030002 & 18.6688479067 & 1.7440050978 & TRES\\
2458143.9292529998 & 14.2423325044 & 1.5732199566 & TRES\\
2458144.8402519999 & -8.9227460931 & 5.1068396065 & TRES\\
2458145.7332580001 & 8.1456954950 & 1.4028823422 & TRES\\
\hline\hline
\end{tabular}

  }
\end{table*}

\subsubsection{WiFeS}
 Four spectra of \thisstar\ were taken using the Wide Field Spectrograph (WiFeS; \citealt{2010ApSS.327..245D}) on the Australian National University's (ANU) 2.3m telescope at Siding Spring Observatory, Australia. WiFeS is an image slicer integral field spectrograph, with a spatial resolution of 1\arcsec per spatial pixel in the $2\times 2$ bin mode. Our observing strategy, reduction, and analyses techniques are detailed in full in (\citealt{Bayliss:2013, Zhou:2016k17}). The observations revealed an early-type star with rapid rotation. No radial velocity variation above $10$\,\kms was noted, but the constraints on the stellar properties and radial velocity were poor due to the rapid rotation of the host star.  

\subsubsection{TRES}
We obtained 11 \emph{R} = 44000 spectra and RVs of \thisstar\ with the Tillinghast Reflector Echelle Spectrograph (TRES; \citealt{Furesz:2008}) on the 1.5m Tillinghast Reflector at the Fred L. Whipple Observatory (FLWO) on Mt. Hopkins, Arizona. The first spectrum -- a reconnaissance spectrum -- was taken on UT 2015 Mar 4, with a 150-s exposure that achieved a signal-to-noise ratio of $49.1$ per resolution element over the Mg b lines; the other 10 spectra, listed in Table \ref{tab:RVs}, were taken between UT 2018 Jan 17-27 with 1200 s to 3000 s exposures, reaching signal-to-noise ratios of $\sim 120$ per resolution element over Mg b. We reduced the spectra following \citet{Buchave:2010}. Radial velocities were derived from these observations by modeling the stellar line profiles with a least-squares deconvolution (LSD) analysis \citep{1997MNRAS.291..658D}. We found that modeling the LSD stellar line profile yielded more reliable radial velocities than other cross correlation techniques for rapidly rotating stars. 

Spectroscopic transits of \thisstar B were also obtained on 2016 Mar 1 and 2016 Mar 19 (UT), with 26 and 24 exposures, respectively, on the two nights. However, the decreased line profile strength due to rapid rotation led to null detections of the spectroscopic transit. We excluded these data from our global analysis.

\subsubsection{McDonald Observatory}

We obtained 14 spectra of \thisstar\ using the Robert G.\ Tull Echelle Spectrograph \citep{Tull:1995} on the 2.7 m Harlan J.\ Smith Telescope at McDonald Observatory on Mt. Locke, Texas. We obtained the spectra using the TS23 spectrograph configuration, giving a resolving power of $R=60000$ over 3570 to 10200 \AA. We obtained the observations between UT 2016 Oct 24 and 2018 Jan 25. The first two spectra had exposure times of $\sim160$ s, while the remainder used 1200~s exposure times, achieving a SNR per resolution element of $\sim 300-560$. We reduced and extracted these spectra using standard IRAF packages, and measured radial velocities using the same methodology as described above for our TRES spectra.

\subsubsection{Spectroscopic transit with Magellan-MIKE}
The spectroscopic transit of \thisstar B was observed with the Magellan Inamori Kyocera Echelle (MIKE; \citealt{2003SPIE.4841.1694B}) on the 6.5\,m Magellan Clay telescope at Las Campanas Observatory, Chile. The series of observations was obtained on 2017 Dec 30, covering the predicted transit itself and baselines prior to ingress and after egress. The observing procedure largely follows that described in \citet{2018AJ....156..250Y}. We made use of the $0\farcs35$ slit, yielding a spectral resolution of $R=80000$ in the blue arm over $3200-5000$\,\AA, and $R = 65000$ in the red arm over $4900-10000$\,\AA. Each exposure had an integration time of 300s. Thorium-Argon arc lamp exposures were obtained every 30\,minutes to provide the wavelength solution. The spectra were reduced and extracted using the Carnegie \emph{Carpy} pipeline \citep{2000ApJ...531..159K,2003PASP..115..688K}. 

\begin{figure}
\includegraphics[width=1\linewidth]{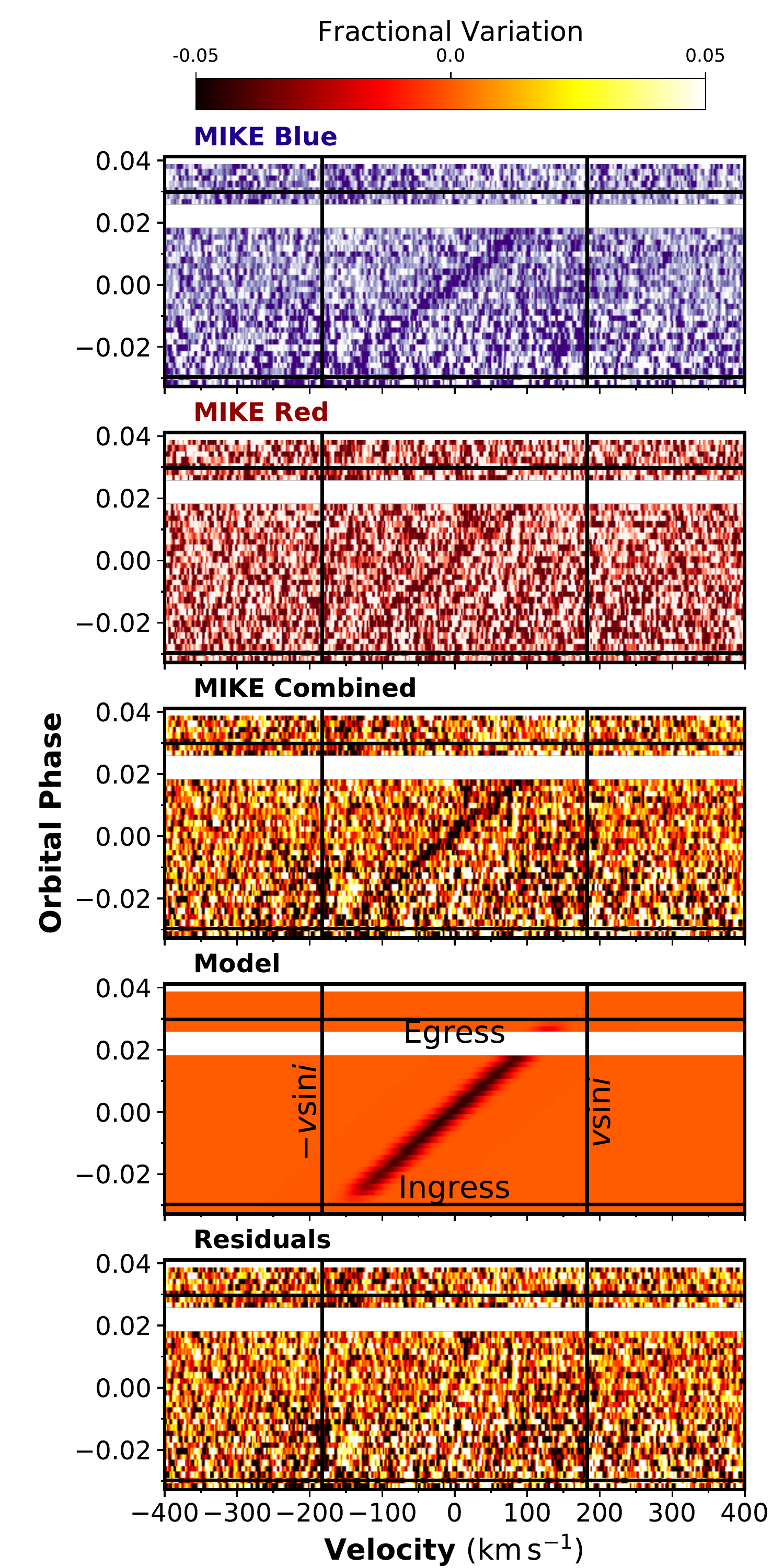}
 \vspace{-.2in}

\caption{Spectroscopic transit of \thisstar\ with MIKE on the Magellan telescope. {\it From top:} Blue-arm, red-arm, and combined observations of the transit; the best-fit spectroscopic transit model; and the residuals. The transit is visible in the MIKE data and absent in the residuals. }
\label{fig:DT} 
\end{figure}

During the transit, the companion star sequentially blocks parts of the rotating surface of the host star. The transit manifests as an indentation on the rotationally broadened line profile of the host star \citep{Collier:2010b}. To reveal this Doppler shadow, we used the spectroscopic time series from MIKE to derive an LSD line profile from each spectrum \citep[following ][]{Zhou:2019}. The mean line profile is then removed from each observation, revealing the shadow cast by the planet. The spectroscopic transit is shown in Figure \ref{fig:DT}, from the red and blue arms independently, as well as from the combined dataset.

\begin{figure}
\includegraphics[width=1\linewidth]{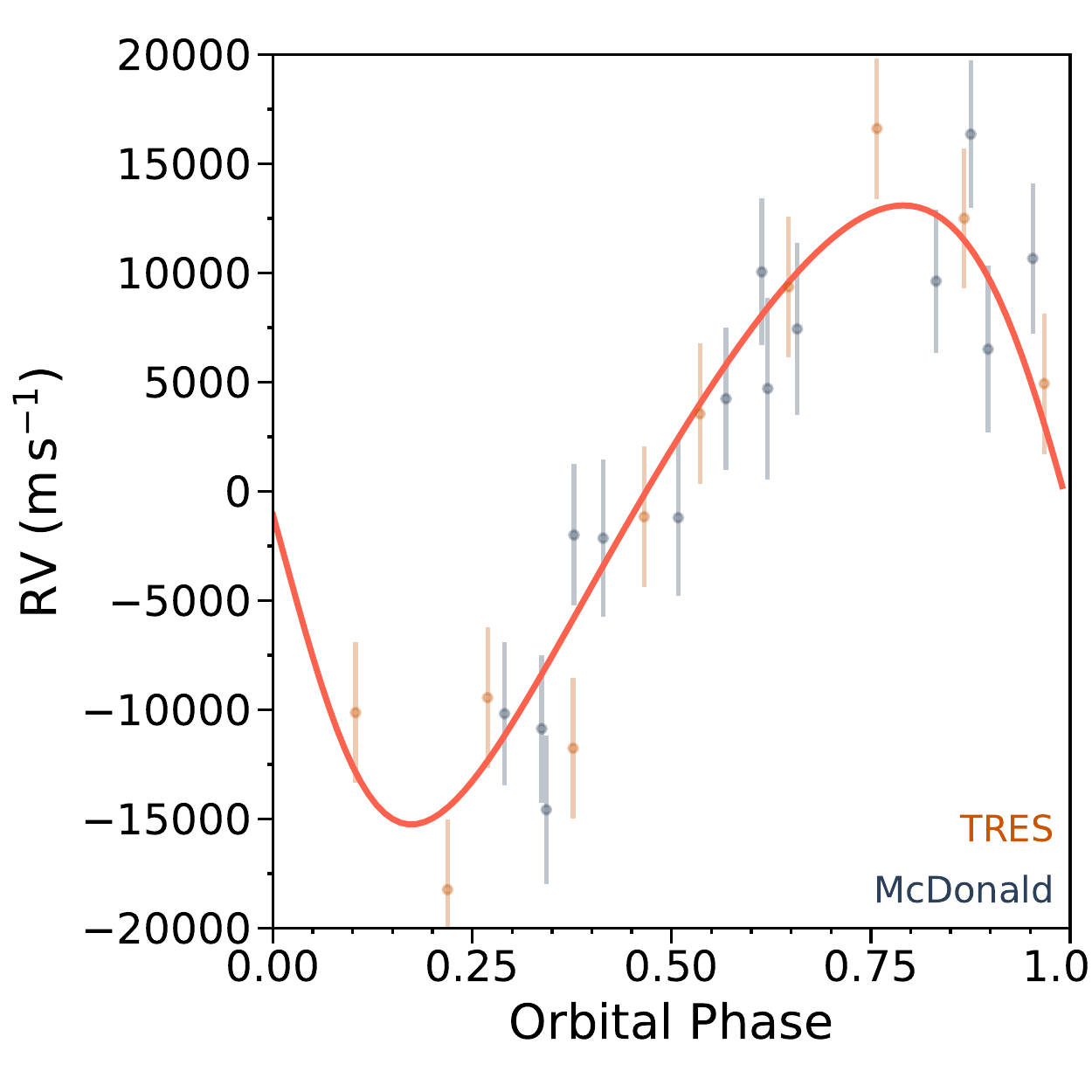}
 \vspace{-.2in}

\caption{TRES (orange points) and McDonald (blue points) radial velocities of \thisstar, along with the best-fit Keplerian RV orbit from Section \ref{sec:GlobalFit} (red-orange line). The model and data are phase-folded to the best-fit orbital period, where phase 0.0 corresponds to mid-transit.  Note that we exclude the MIKE RV data and the near- and in-transit TRES data from our analysis and thus do not attempt to model the Rossiter-McLaughlin effect; rather, we model the Doppler Tomography signal from the spectra taken by MIKE before, during, and after transit.}
\label{fig:RV} 
\end{figure}

\subsection{High Resolution Imaging Follow-Up}
\subsubsection{Keck Observatory}

In order to search for any potential third stellar object in the system, or any background stars that could contaminate the photometry or spectroscopy, we observed \thisstar\ with the NIRC2 imager on the Keck II Telescope, Maunakea, Hawaii, on UT 2018 Jan 4. 
We observed in two different modes: conventional high-contrast AO imaging, and non- redundant aperture masking interferometry \citep[NRM;][]{tuthill2000,krilandlkca15,dmys1}. We took the observations in the $K’$ filter in the vertical angle mode without dithering. In the latter mode, we introduced the nine-hole mask into the pupil plane. For all observations, we used the smallest possible pixel scale for NIRC2 (9.952 mas; \citealt{Yelda2010,Yelda2011}), and we observed two nearby calibrator stars chosen both for similarity in Gaia colours and K-band magnitude and for their proximity to \thisstar\ ($<$10$^\circ$ separation) on the sky.

For the conventional imaging observations, we took six exposures, each with 40 co-adds of 0.5 s and two Fowler samples. For the NRM observations, we resampled the telescope into a sparse interferometric array by placing a mask into the pupil plane of the telescope. This allows the use of the complex triple-product, or closure-phase observable, to remove non-common path errors produced by atmospheric conditions, or variable optical aberrations. We obtained eight interferograms, each with a single 20-s coadd and 64 Fowler samples. For the two calibrator stars (KELT-19 and BD+12 1601), we obtained one and three images, respectively; we also obtained eight interferograms of each, all in identical setups.

Our conventional-imaging data reduction and analysis follows the description given in \citet{Kraus2016}. We employed frame-by-frame point-spread function (PSF) subtraction using two methodologies. For faint, wide companions beyond 500 mas, we subtracted an azimuthally averaged flux profile of \thisstar. To probe closer to the inner working angle and reduce speckle noise, we subtracted a scaled, best-fitting empirical PSF produced using the calibrator star images. We created significance maps for each image by measuring flux in 40 mas apertures entered on each image pixel. These maps where then stacked (weighted by Strehl ratio) to compute a final significance map centered around \thisstar. We measure detection limits as a function of angular separation from the primary by inspecting the distribution of confidences in 5-pixel annuli. Neither \thisstar\ nor the calibrator stars exhibit a statistically significant flux excess within the NIRC2 field of view; any pixel with $>6\sigma$ total confidence would have been considered a candidate companion and inspected further to confirm that it was not a residual speckle, cosmic ray, or image artifact.

 Our NRM data reduction and analysis follows the procedures of \citet{Kraus2008} and \citet{Ireland2013}. We used observations of the two calibrators to remove systematics in the closure-phase observable. We then fit the calibrated closure phases with binary source models to search for significant evidence of a companion, and we calibrated detection limits using a Monte-Carlo process of randomizing phase errors and determining the distribution of possible binary fits. As with the conventional imaging, we detected no significant sources (aside from \thisstar) in the masking data for \thisstar.

\begin{figure}
\includegraphics[width=1\linewidth]{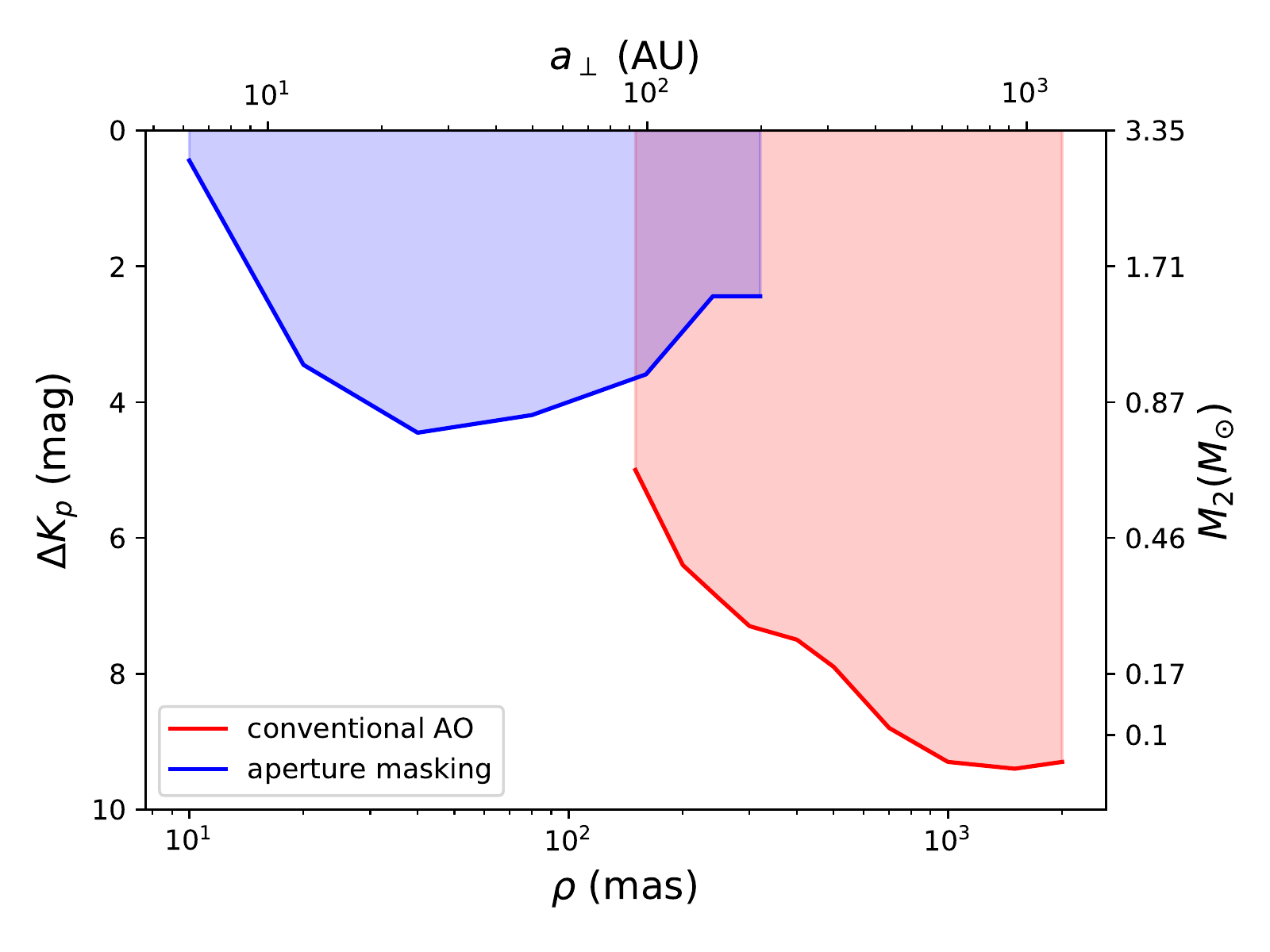}
\caption{$5\sigma$ contrast curves from our NIRC2 imaging, with the conventional AO and coronagraphic curve in red and the non-redundant aperture masking interferometric curve in blue. The excluded regions are shaded. The upper axis shows the corresponding projected separation, given the {\it Gaia} DR2 distance to the system. The right axis shows the approximate secondary mass corresponding to each $\Delta K_p$ value, estimated using the \texttt{isochrones} code \citep{isochrones}. The MIST isochrones (\citealt{Dotter16, Choi16}) used by \texttt{isochrones} only go down to $0.1 \MS$, close to the hydrogen-burning limit below which the magnitude of an object is a strong function of its age. We therefore do not plot any tick marks below the $0.1 \MS$ level.}
\label{fig:AO} 
\end{figure}

We show our derived contrast curve in Fig.~\ref{fig:AO}. Using the NRM data, we can exclude stellar companions as faint as $\Delta K_p\sim4$ with separations of a few tens of mas (corresponding to projected physical separations of $\sim20-40$ au at the distance of \thisstar). Our conventional AO imaging, meanwhile, allows us to exclude companions with $\Delta K_p<5$ at 0\farcs15 (90 au), with limits improving to $\sim9$ magnitudes at 2\farcs0 ($\sim1000$ au). Note that, while we would have detected a star like \thisstar B if there were one at large projected separations ($\gtrsim 300$ au), neither high-contrast imaging method would reveal an M star at shorter separations. This highlights the difficulty of detecting binary systems like \thisstar\ with high-contrast imaging.

\subsubsection{Gaia DR2 Comoving Companion Search}

As we have found that \thisstar\ is young (\S\ref{sec:GlobalFit}), with an age of $183$ Myr, it is possible that it could still be associated with other stars that formed with it as part of a moving group. We used {\it Gaia} DR2 to search for potential comoving companions outside the field of view of our Keck imaging observations. There are 20,654 {\it Gaia} DR2 sources within a projected separation of 10 pc (55.9') of \thisstar. Of these, 160 have proper motions differing from those of \thisstar\ by less than 1$\sigma$, and either a parallax differing by less than $1\sigma$ or distances of less than 10 pc from \thisstar. All of these comoving candidates, however, are very faint and have parallax and proper motion uncertainties of $>0.5$ mas and $>0.9$ mas yr$^{-1}$, respectively. We thus conclude that these are likely to be spurious matches simply due to low-quality measurements of these faint sources.  Additionally, 222 of the 20,654 {\it Gaia} DR2 sources have radial velocity measurements. None of these differ from the absolute RV of \thisstar\ by less than 5$\sigma$, and also have parallaxes and proper motions disagreeing by less than 5$\sigma$. We thus conclude that there is no evidence for any comoving companions to \thisstar\ within 10 pc down to a magnitude difference of $\Delta G\sim12$, corresponding to a candidate mass of approximately 0.17 \MS\ \citep[using the \texttt{isochrones} code;][]{isochrones}.

\section{Analysis and Results}
\subsection{Global Fit Results}
\label{sec:GlobalFit}

Modeling the \thisstar\ system incorporates a series of factors not usually accounted for in traditional transiting exoplanet models. 

Rapidly rotating stars are oblate in shape, with a smaller polar radius than equatorial radius by up to $\sim 10\%$. This deviation from the spherical geometry induces gravity darkening in the stellar surface brightness, with the polar surface brightness being brighter and hotter than the equatorial surface brightness \citep{1924MNRAS..84..684V}. To account for the effects associated with a rapidly rotating primary star, our modeling follows the process laid out in \citet{Zhou:2019}, and differs from standard planetary and binary stellar models in the following ways:
\begin{itemize}
    \item We utilize a set of disk-integrated SEDs that accounts for the gravity darkening effect as viewed from different line-of-sight inclinations $(I*)$. The same rotating star appears cooler and fainter when viewed equator on, and hotter and brighter when viewed pole-on. At each iteration of the model fitting, we interpolate our SED grid against the tested stellar mass $M_1$, radius $R_1$, metallicity [Fe/H], rotational velocity $\vsini$, $I_\star$, parallax, and interstellar reddening $E(B-V)$ of the system, and compute the $\chi^2$ difference between the interpolated SED and the observed APASS and 2MASS results. 
    \item We account for the deformed shape of the transit, as the stellar rotation deforms the primary star into an oblate shape. For an oblate primary star, the transit duration would be longer along the equator, with projected obliquity of $|\lambda| = 0^\circ$, and shorter from pole to pole, with $|\lambda| = 90^\circ$. 
    \item The light curve of a transit across a gravity-darkened star is dependent on the obliquity angle \citep{Barnes2009}. We adopt the \texttt{simultrans} package \citep{2018AJ....155...13H} to account for the transit gravity darkening effect in our light curve modeling.
    \item The primary star stellar parameters are constrained by the Geneva rotational isochrones \citep{2012A&A...537A.146E}. Importantly, the stellar oblateness is inferred from the isochrone at each iteration based on the trial primary stellar mass $M_1$, radius $R_1$, metallicity [Fe/H], rotational velocity $\vsini$, and $I_\star$.
\end{itemize}

The non-negligible mass and luminosity of the secondary star also induces phase variations in the light curve. We incorporate the phase variations of the system via a BEaming, Ellipsoidal and Reflection/thermal variation (BEER) analysis \citep[e.g][]{Faigler:2011,Esteves:2013}. We decompose the phase variations to that contributed by the star's phase function $F_\mathrm{p}$, secondary eclipse $F_\mathrm{ecl}$, Doppler boosting $F_\mathrm{d}$, and ellipsoidal variations $F_\mathrm{e}$, following equation 1 and associated subsequent formulations laid out in \citet{Esteves:2013}. We fit for the amplitudes of each component simultaneous with the global modeling.

We use the available discovery and multi-band follow-up light curves, RVs, DT measurements, Gaia DR2 parallax \citep{2018A&A...616A...1G}, APASS \citep{Henden:2015} and 2MASS \citep{Skrutskie:2006} magnitudes in our global modeling. We model the entire dataset with free parameters describing the orbital period $P$, transit ephemeris $T_0$, mass of primary star $M_1$, mass of secondary star $M_2$, radius of primary star $R_1$, line of sight inclination $I_\star$, metallicity [Fe/H], parallax, and interstellar reddening $E(B-V)$. Parameters that largely govern the photometric transit include the inclination $i$, radius ratio $R_2/R_1$, and gravity darkening coefficient $\beta$, as per \citet{1924MNRAS..84..684V}. RV parameters include $\sqrt{e} \cos \omega$, $\sqrt{e} \sin \omega$, velocity offsets $\gamma$ for each instrument, and velocity jitter parameters for each instrument. The spectroscopic transit requires the additional parameters $\lambda$, $v\sin I_\star$, and the velocity of the non-rotational stellar broadening component $v_\mathrm{broad}$; the latter is modeled as per Gray's radial tangential function \citep{Gray1976}. 

As in \citet{Zhou:2019}, we sample the parameters' posteriors with the Markov Chain Monte Carlo method implemented in \texttt{emcee} \citep{Foreman-Mackey2013}. We adopt Gaussian priors on the rotational broadening velocity, based on its spectroscopically measured value; [Fe/H], based on the Galactic disk metallicity at $0.15-1.00$ Gyr \citep{Robin2003}; and the system parallax, using the {\it Gaia}\ DR2 value \citep{Gaia:2018} with the \citet{Stassun:2018} systematic correction. We place upper limits on the allowed interstellar reddening values, corresponding to the maximum reddening value from \citet{Schlafly2011}, and on $V_{\rm non-rot}$, corresponding to the width of the planetary signal in the DT data. We place uniform priors on $\cos I_\star$, between 0 and 1, and $\beta$, between 0 and 0.3 as motivated by theoretical models of gravity darkening (see e.g. \citealt{Espinosa2011}).

\begin{figure}
\includegraphics[width=1\linewidth]{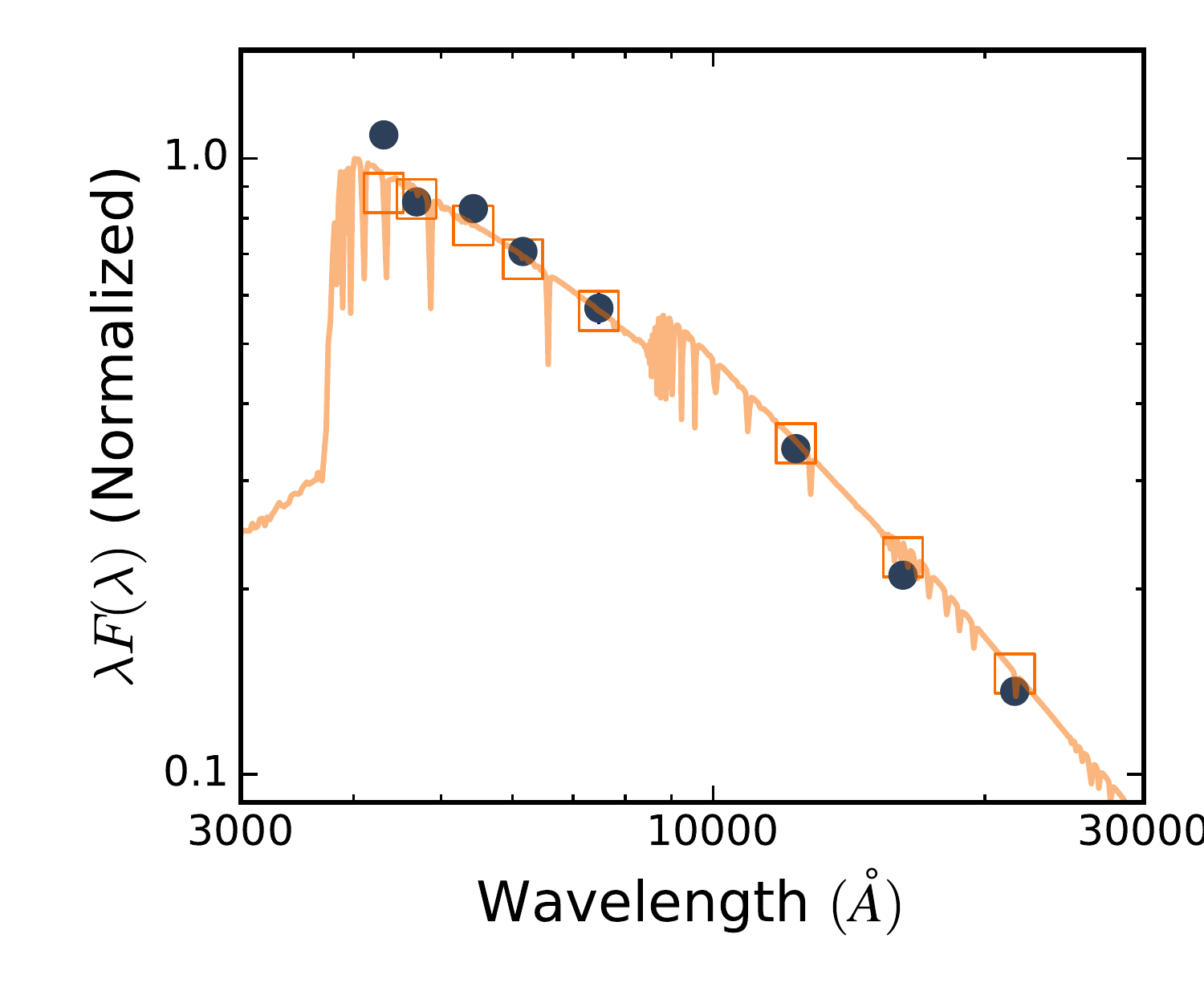}
 \vspace{-.2in}  \label{fig:SED}

\caption{Model spectral energy distribution (SED; orange lines) of \thisstar. The literature broad-band photometric measurements are the black points, while the model-predicted values in each corresponding passband are the red squares.}
\end{figure}

Table \ref{tbl:parameters} lists the system parameters from this analysis, while the data and best-fit models are shown in Figures \ref{fig:All_light curve},\ref{fig:tess} \ref{fig:RV}, \ref{fig:DT}, and \ref{fig:SED}. We find that \thisstar A has a mass of $3.34\MS$, a polar radius of $3.1\RS$, a slight oblateness quantified by a polar-to-equatorial radius ratio of 0.95, an effective temperature of $12,000$ K, and a luminosity of $180\LS$; these parameters are consistent with a B9V spectral type \citep{Pecaut:2013}.

We also find \thisstar B to have a mass of $0.22\MS$ and a radius of $0.30\RS$, with a surface brightness in the TESS bandpass that is $4\%$ of \thisstar A's at the time of secondary eclipse. We infer that the system is young ($183$ Myr) which suggests that \thisstar B may not yet be on the Main Sequence -- see Section \ref{sec:inflation} for a comparison to low-mass isochrones. Despite the young age, for which one would not yet expect tidal forces to produce spin-orbit alignment, we infer that the orbit of \thisstar\ is aligned, with a small projected obliquity, $\lambda = 1.3\degr$. 

We note that visible discrepancies between the best-fit model light curve and the data -- such as the later-than-observed model ingress of the TESS data -- may result from any uncorrected systematic effects in the data themselves or from any missing physics in our model, including our simple treatments of tidal deformation and gravity darkening.

\begin{table*}
\centering
\setlength\tabcolsep{1.5pt}
\caption{Median values and 68\% confidence interval for the physical and orbital parameters of the \thisstar\ system}
  \label{tbl:parameters}
  \begin{tabular}{llrr}
  \hline
  \hline
  Parameter & Description (Units) & Value & Priors  \\
  & & \\
 \hline
Fitted Parameters: & & &\\
\hline
{\it Stellar Parameters:} & & &\\
$M_1$ & Primary stellar mass (\msun) & $3.34_{-0.09}^{+0.07}$ & \\
$R_1$ & Primary stellar radius (\rsun) & $3.10_{-0.10}^{+0.08}$ & \\
$M_2$ & Companion stellar mass (\msun) & $0.22 \pm 0.02$ & \\
$R_2/R_1$ & Radius ratio & $0.0977 \pm 0.0008$ &\\
$\rm [Fe/H]$ & Metallicity & $-0.01_{-0.1}^{+0.1}$ & $\mathcal{G} (0.03,0.13)^{a}$\\
$\beta$ & Gravity darkening parameter & $0.049_{-0.029}^{+0.048}$ & $\mathcal{U}(0,0.3)^{b}$\\
$V_{\rm non-rot}$ & Non-rotational line broadening (km/s) & $12 \pm 1$ &$\mathcal{U}(0,50)$ \\
$V\sin I_{*,1}$ & Rotational broadening (km/s) & $183 \pm 1$ & $\mathcal{G}(183.239,1.46)$\\
\\
{\it Orbital Parameters:} & & &\\
$T_0$ & Time of primary eclipse center $({\rm BJD_{TDB}})$ & $2457064.72887_{-0.00090}^{+0.00089}$ &\\
$P$ & Orbital period (days) & $3.62187347_{-2.64\times 10^{-6}}^{+2.91\times 10^{-6}}$ & \\
$\sqrt{e} \cos \omega$ & -- & $0.010_{-0.0275}^{+0.0396}$ & \\
$\sqrt{e} \sin \omega$ & -- & $0.114_{-0.189}^{+0.273}$ &\\
$i$ & Inclination (deg) & $95.27_{-0.88}^{+0.69}$ & \\
$I_{*,1}$ & Rotation axis inclination (deg) & $78\pm 8$ &$\cos I_\star^{c}(0,1)$ \\
$\lambda$ & Projected spin-orbit alignment (deg) & $1.32_{-0.71}^{+0.78}$ &\\
$\pi$ & Parallax (mas) & $1.69\pm 0.05$ & $\mathcal{G} (1.6250,0.0785)$ \\
$E_{B-V}$ & Reddening & $0.016_{-0.015}^{+0.016}$ & $\mathcal{U}(0,0.0415)$\\
\\
{\it RV \& Phase\ Curve\ Parameters:} & & & \\
$A_{\rm refl}$ & Reflection amplitude (ppm) & $415_{-92}^{+91}$ & \\ 
$A_{\rm beam}$ & Doppler beaming amplitude (ppm) & $29_{-35}^{+37}$ & \\ 
$A_{\rm ellip}$ & Ellipsoidal variation amplitude (ppm) & $652_{-46}^{+43} $ & \\
$H_{\rm dilute}$ & Dilution factor for KELT &$0.776_{-0.038}^{+0.039}$\\ 
$\gamma_{\rm McDonald}$ & RV offset (km/s) &$2_{-1}^{+1}$ &\\
$\gamma_{\rm TRES}$ & RV offset (km/s) & $10 \pm 1$ &\\
$\sigma_{\rm McDonald}$ & RV jitter (km/s) & $3 \pm 1$ &\\
$\sigma_{\rm TRES}$ & RV jitter (km/s) & $3 \pm 1$ &\\
$\sigma_{\rm Gaia}$ &{\it Gaia} photometry jitter (ppm)& $886_{-664}^{+2428}$ & \\
\\

Derived Parameters: & & &\\
\hline
{\it Stellar Parameters:} & & &\\
$R_{1, \rm pole}/R_{1, \rm eq}$ & Primary stellar oblateness & $0.935_{-0.002}^{+0.004}$&\\
$T_{\rm eff,1}$ & Primary effective temperature (K) & $11960_{-520}^{+430}$ & \\
$\log g_1$ & Primary surface gravity & $4.01 \pm 0.04 $ & \\
$L_1$ & Primary luminosity (\lsun) & $179_{-28}^{+26}$& \\
${\rm Age}$ & Primary stellar age (Myr) & $183_{-30}^{+33}$&\\
$V_{\rm crit}$ & Breakup velocity (km/s) & $451 \pm 8$ & \\
$V/V_{\rm crit}$ & Rotation-to-breakup velocity ratio & $0.416 \pm  0.009$ &  \\
$S_2/S_1$ & Surface brightness ratio & $0.052_{-0.016}^{+0.015}$ & \\
$R_2$ & Companion stellar radius (\rsun) & $0.303_{-0.010}^{+0.007}$ & \\
\hline
{\it Orbital Parameters:} & & & \\
$e$ & Eccentricity & $ < 0.07\ (1\sigma)$ & \\
$a$ & Semi-major axis (a.u.) & $0.0689_{-0.0006}^{+0.0004}$ & \\
$a/R_1$ & Semi-major axis, in stellar radii & $4.77_{-0.10}^{+0.16}$ & \\
$b$ & Impact parameter & $0.44_{-0.06}^{+0.05}$& \\
$T_{14} $ & Primary eclipse duration (days) & $0.246_{-0.005}^{+0.003}$ & \\
$K_{\rm RV}$ & RV semi-amplitude (km/s) & $13.89_{-1.31}^{+1.41}$ & \\
 \hline
 \hline
 \end{tabular}
\begin{flushleft}
 \footnotesize \textbf{\textsc{NOTES}} \\
  \vspace{.1in}
  \footnotesize $^a$ Gaussian prior (mean, standard deviation).\\
   \footnotesize $^{b}$ Uniform prior (lower limit, upper limit).\\
 \end{flushleft}
\end{table*}

\subsection{\label{sec:uvw}Location and Motion in the Galaxy}

We determine the motion of \thisstar\ through the Galaxy to place it in one of the Galactic stellar populations. We adopt an absolute RV of $11.15\pm 0.13 \rm{km\ s^{-1}}$ by adding together the binary's barycentric RV ($\gamma_{\rm TRES}$ in Table \ref{tbl:parameters}), the offset between the RVs recovered from the least-squares deconvolution and cross-correlation (CCF) methods, $1.755\ \rm{km\ s^{-1}}$, and the known offset of $-0.610\ \rm{km\ s^{-1}}$ between the TRES RVs and the IAU standard scale. We list the individual relative RVs and their uncertainties in Table \ref{tab:RVs}. We calculate {\it U, V,} and {\it W} space velocities by combining the adopted absolute RV with a parallax and proper motions from {\it Gaia} DR2 (\citealt{Gaia:2018, Lindegren:2018}), with the parallax corrected for the $0.0820~\mu$as systematic offset found by \citet{Stassun:2018}.  We adopt the \citet{Coskunoglu2011} solar velocity with respect to the Local Standard of Rest. We note that we could have used the estimate of the distance to \thisstar\ from the global analysis (see \ref{tbl:parameters}).  However, given that the global analysis does not provide an absolute RV or a proper motion of the star, we chose to just use extrinsically-determined quantities for this analysis.  We note that the GAIA DR2-derived distance and the distance derived by the global fit differ by only $\sim 0.8\sigma$.

For \thisstar, $(U,V,W) = (-3.16\pm 0.25, 8.08 \pm 0.32, -0.32 \pm 0.58)$ -- all in units of $\rm{km\ s^{-1}}$ -- where positive $U$ points toward the Galactic Center. We find a $99.5\%$ probability that \thisstar\ belongs to the thin disk, according to \citet{Bensby:2003}.

Furthermore, we find that \thisstar\ is located $615\pm 30$~pc away from the Sun.  At $b\simeq 11^\circ$, this system has a vertical (Z) distance from the Sun of $Z-Z_\odot=120$pc.  Taking into account the \citet{Bovy:2018} distance of the sun above the Galactic plane of $Z_\odot \simeq 30$~pc as determined from giant stars, this implies a vertical distance of this star from the Galactic plane of $\sim 150$~pc. This is a surprisingly large $Z$ distance, given the the scale height of late B/early A stars is $\sim 50$~pc \citep{Bovy:2018}.

\begin{figure}
    \centering
    \includegraphics[width=1.\linewidth,trim=0 0 0 50,clip]{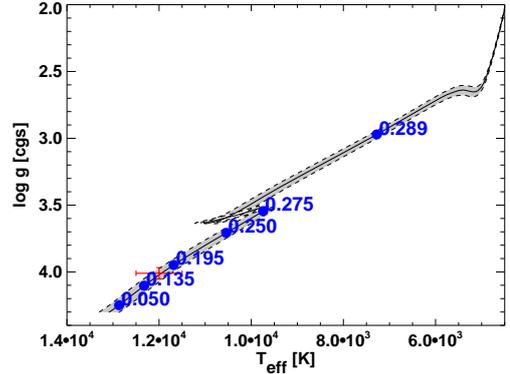}
    \caption{Kiel diagram showing the evolutionary track (solid line) and $1\sigma$ contours (dashed lines; grey shading) for \thisstar\ A (red cross) using the best-fit stellar parameters from Section \ref{sec:GlobalFit}. Blue points denote locations along the track at the specified ages. \thisstar\ A is a little more than halfway to the end of its main-sequence lifetime.}
    \label{fig:hrd}
\end{figure}

To show the inferred evolution of \thisstar A, we plot a Kiel $\logg-\teff$ diagram in Figure \ref{fig:hrd}. Given our inferred age of 183 Myr, \thisstar\ A is likely not an evolved star. The age of \thisstar A, its Galactic space velocities, and its location on a Kiel diagram are all consistent with the inference that this system is relatively young.  However, the relatively large distance of \thisstar\ above the plane is somewhat surprising.  

\subsection{BEER Analysis of Out-of-eclipse Variations}\label{subsec:beer}
As a point of comparison for the analytic out-of-eclipse modeling performed as part of the global fit in Section \ref{sec:GlobalFit}, we adapt the equations of \citet{Shporer2017} to perform a separate BEER analysis \citep{Faigler:2011} of the out-of-eclipse TESS photometry.
\begin{enumerate}
    \item {\bf Doppler Beaming:} The beaming effect is analogous to RV blue- and red-shifts of stellar spectral lines due to the gravitational influences of nearby companions. This effect leads to the flux emitted along our line of sight being shifted to higher or lower energies, depending on the phase of the orbit, which produces flux variations that depend on the wavelength range being observed and the radial velocity and SED of the emitting body. In terms of the RV semiamplitude $K$ and the speed of light $c$, the relative amplitude of the beaming effect is given by $\alpha_{\rm beam}\frac{4K}{c}$, where $\alpha_{\rm beam}$ is an order-unity function of the star's rest-frame SED slope within the observed wavelength range weighted by the transmission curve of the corresponding bandpass (see  \citealt{Loeb:2003}, \citealt{Shporer2017}). In terms of stellar and orbital parameters, this becomes
    \begin{multline}
            A_{\rm beam} = 0.0028\ \alpha_{\rm beam}\left(\frac{P}{\rm day}\right)^{-1/3}\\ \times \left(\frac{M_1+M_2}{\MS}\right)^{-2/3}\left(\frac{M_2\sin i}{\MS}\right)
    \end{multline}
    \item {\bf Ellipsoidal Variations:} A massive body (e.g. star or planet) can also distort the shape of its binary companion, producing ellipsoidal variations. To first order in the equilibrium-tide approximation (valid for stars with thick convective envelopes),
    \begin{multline}
         A_{\rm ell} = \frac{64}{5}\alpha_{\rm ell}\sin^2 i\left(\frac{R_1}{\RS}\right)^3 \left(\frac{P}{\rm day}\right)^{-2}\left(\frac{M_2}{\MJ}\right)\left(\frac{M_1+M_2}{\MS}\right)^{-2},
    \end{multline}
    where, for a primary star linear limb-darkening coefficient $u_1$ and gravity-darkening coefficient $\beta_1$,
    \begin{equation}
        \label{eq:ell}
        \alpha_{\rm ell} = 0.15\frac{(15+u_1)(1+\beta_1)}{3-u_1}
    \end{equation}
    \item {\bf Reflected and Re-radiated Light}: Finally, additional flux variations can be caused by one body reflecting, scattering, or absorbing and re-radiating the incident flux from the other. The combination of these effects is modeled as
    \begin{multline}
        \label{eq:refl}
        A_{\rm ref} = 57\alpha_{\rm ref}\sin i\left(\frac{M_1+M_2}{\MS}\right)^{-2/3}\left(\frac{P}{\rm day}\right)^{-4/3}\left(\frac{R_2}{\RJ}\right)^2.
    \end{multline}
    Here, $\alpha_{\rm ref} = 10p_{1, \rm geo,eff}$, where $p_{1,\rm geo,eff}$ is the effective geometric albedo accounting for reflected and scattered light in the observed bandpass and any phase-dependent re-radiation of absorbed incident flux. We note that our use of the term ``albedo'' is a misnomer: our albedo incorporates any contributions from scattered and/or re-radiated light in addition to reflected light, though the stars' intrinsic fluxes in the TESS bandpass are modeled separately.
\end{enumerate}

Expressions for the companion star can be obtained by swapping the subscripts 1 and 2 in the above equations. These equations are similar to those in, e.g. \citet{Shporer2017}, but modified to account for the non-negligible mass of a stellar companion.

We fit the out-of-eclipse data for the stellar parameters $M_1$, $R_1$, $T_{\rm eff,1}$, $M_2$, $R_2$, $T_{\rm eff,2}$; the orbital inclination $i$, and  a zero-point offset, $a_0$. We use the global fit results as starting values for these parameters and enforce Gaussian priors on $M_1$, $R_1$, $T_{\rm eff,1}$, and $i$ with widths equal to the average of the upper and lower $1\sigma$ uncertainties from the global fit. We also adopt a prior on the surface brightness ratio, $S_2/S_1$, which is important for the beaming calculation as described below.

For \thisstar A, we fix $\beta_1$ to the best-fit value from the global fit. We adopt the other gravity-darkening and linear limb-darkening coefficients from \citet{Claret:2017}. We use the best-fit $\logg$, $\feh$, and $T_{\rm eff,1}$ to determine a linear limb-darkening coefficient of $u_1 = 0.295$; we adopt a coefficient for the companion, $u_2 = 0.5$, based on the $\logg$ calculated from the global fit and assuming that 1) the companion's metallicity is the same as the primary star's metallicity and 2) the companion is cool, with $T_{\rm eff,2} \sim 3200$ K as inferred from the 200 Myr isochrone used in Section \ref{sec:inflation}. We use the same input parameters to obtain $\beta_2$.

We adopt an albedo of 1 for the primary and 0.5 for the companion based on calculations from \citet{Rucinski1969} and \citet{Rucinski:1989}. We calculate $\alpha_{\rm beam}$ following \citet{Loeb:2003,Shporer2017} and using the TESS filter's response function and NextGen model stellar atmospheres \citep{Hauschildt:1999}. We use the trial values of each stars' effective temperature, with the fixed surface gravity and metallicity values as described in the previous paragraph, to generate an SED for each star. We use the SEDs both to compute $\alpha_{\rm beam}$ and to determine the surface brightness in the TESS bandpass. We use these surface brightnesses to determine the flux ratio in the TESS bandpass and to compare against the prior surface brightness value from the global fit. The surface brightness ratio from the global fit is computed from the secondary eclipse depth and the stellar radius ratio, so it includes the loss of light from both the companion's intrinsic flux and the incident flux it reflects and/or re-radiates. We account for this by adding the reflection amplitude from Equation \ref{eq:refl} to the intrinsic flux ratio, multiplying by $(R_1/R_2)^2$, and comparing against the global fit surface brightness ratio. 

In contrast to phase curve analyses of star-planet systems, \thisstar B may exhibit its own flux variations due to its relatively higher mass and luminosity. For this reason, we also calculate \thisstar B's BEER amplitudes; the ellipsoidal variation signal has the same sign as that for \thisstar A, while the beaming and reflection effects is of opposite sign. The amplitudes reported in Table \ref{tab:beer} are the amplitudes of the combined signals. For this system, the M dwarf's BEER amplitudes are negligible ($<5$ ppm).

\begin{figure}
    \centering
    \includegraphics[width=1\linewidth]{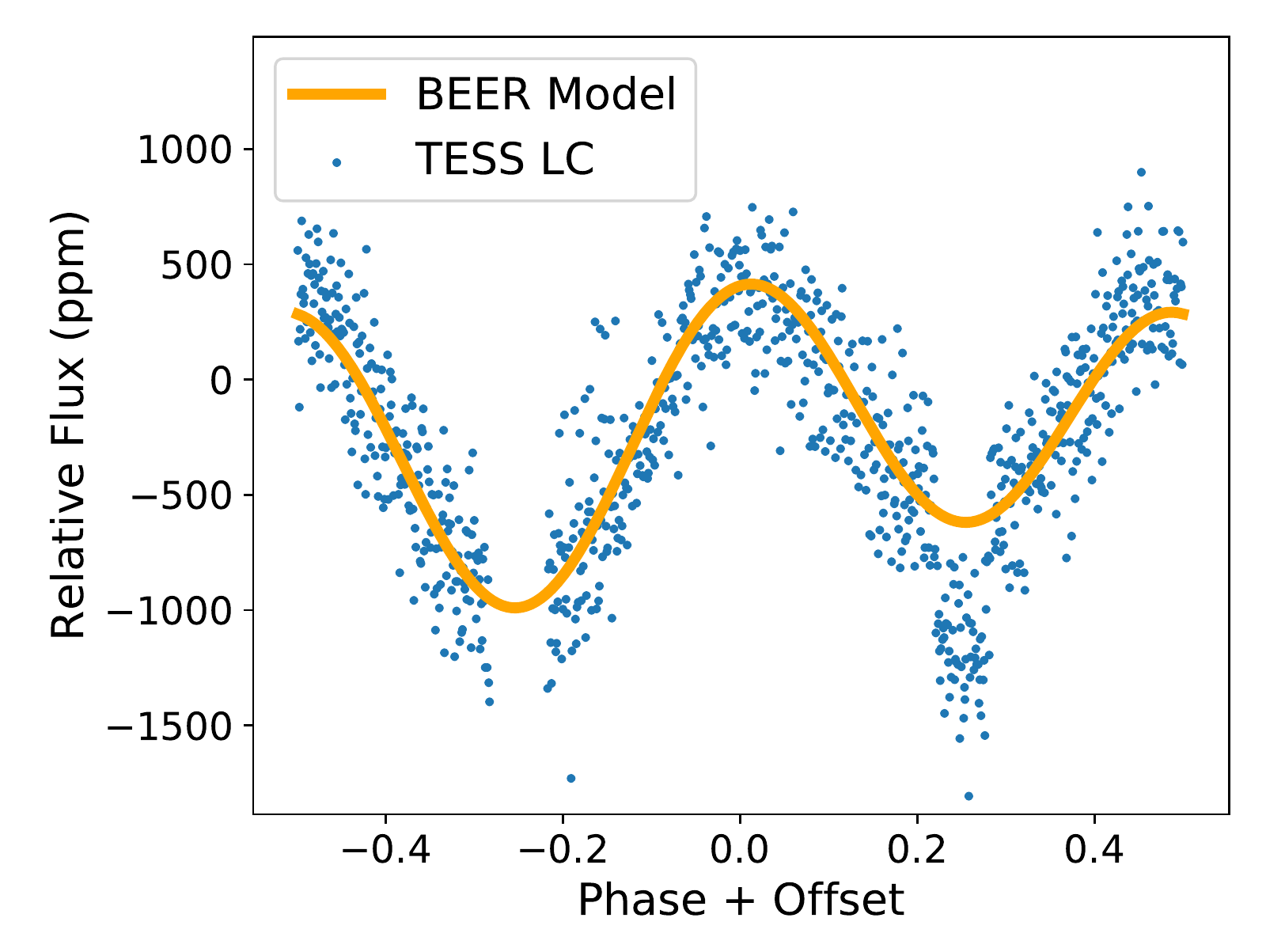}
    \caption{TESS light curve (blue points) and best-fit BEER-only model (orange line). Both are phase-folded to the best-fit ephemeris from Table \ref{tbl:parameters} in Section \ref{sec:GlobalFit}.}
    \label{fig:beer}
\end{figure}

Figure \ref{fig:beer} shows the phase-folded TESS light curve (with eclipses included) and the best-fit BEER model. The ellipsoidal variations are readily apparent, as noted in Section \ref{sec:GlobalFit}, and are well-fit by the model. The difference in model depth between the minima at the primary and secondary eclipses can be explained by the fact that the analytic prescription for the reflected light effect reaches its minimum and maximum at primary and secondary eclipse, respectively.

We find a combined beaming amplitude of $61 \pm 5$ parts-per-million (ppm). We find an effective temperature for \thisstar B of  $T_{\rm eff,2} = 3300_{-1500}^{+600}$ K. While poorly constrained, this temperature is consistent with the isochrone-predicted effective temperature used to compute the companion's limb-darkening coefficient.

We also measure ellipsoidal variation and reflection/reradiation amplitudes of $576 \pm 13$ and $185 \pm 13$, respectively. Despite the two different phase functions used for the reflection effect, we find no significant differences between the inferred parameters. We discuss some caveats to our analysis in Section \ref{sec:disc}, but we leave a more detailed study of the analytic BEER formulae and their application to binaries like \thisstar\ to future work.

\begin{table*}
\centering
\setlength\tabcolsep{3pt}
\caption{\thisstar\ BEER Results: Median Values \& $68\%$ Confidence Interval}
  \label{tab:beer}
  \begin{tabular}{llrr}
  \hline
  \hline
  Parameter & Description (Units) & Value & Priors  \\
 \hline
Fit Parameters & & & \\
~~~$M_{1}$\dotfill &Primary star mass (\MS)\dotfill & $3.27 \pm 0.08$& $\mathcal{G}^{a}$\\
~~~$R_{1}$\dotfill & Primary star radius (\RS)\dotfill & $3.54 \pm 0.08$ & $\mathcal{G}$ \\
~~~$T_{\rm eff,1}$\dotfill &Primary star effective temperature (K)\dotfill & $12200_{-450}^{+500}$ & $\mathcal{G}$ \\
~~~$M_{2}$\dotfill &Companion mass (\MS)\dotfill & $0.17 \pm {0.01}$ & --\\
~~~$R_{2}$\dotfill &Companion radius (\RS)\dotfill & $0.29 \pm 0.01$ & --\\
~~~$T_{\rm eff,2}$\dotfill &Companion effective temperature (K)\dotfill & $3300_{-1500}^{+600}$ & -- \\ 
~~~$i$\dotfill &Inclination (degrees)\dotfill & $95.2 \pm 0.8$ & $\mathcal{G}$\\
~~~$a_0$\dotfill & Flux normalization\dotfill & $0.999771 \pm {0.000008}$ & --\\
Derived Parameters & & & \\
~~~$F_2/F_{\rm tot}$\dotfill &Companion-to-total flux ratio (ppm) \dotfill & $87_{-86}^{+133}$& -- \\
~~~$A_{\rm ell}$\dotfill & Combined ellipsoidal variation amplitude (ppm) & $576 \pm 13$ & -- \\
~~~$A_{\rm ref}$\dotfill & Combined reflection/re-radiation amplitude (ppm4) & $185 \pm 13$ & -- \\
~~~$A_{\rm beam}$\dotfill & Combined Doppler beaming amplitude (ppm) & $61 \pm 5$ & -- \\
 \hline
 \hline
 \end{tabular}
\begin{flushleft}
 \footnotesize \textbf{\textsc{NOTES}} \\
  \vspace{.1in}
  \footnotesize $^a$ Gaussian prior derived from parameters listed in Table \ref{tbl:parameters}.\\
    \footnotesize $^b$ Uniform prior over specified range.\\
 \end{flushleft}
\end{table*}

\subsection{Comparison to Low-mass Stellar Isochrones}\label{sec:inflation}
In Sections \ref{sec:GlobalFit} and \ref{subsec:beer}, we infer a companion mass that is consistent with a fully convective M star. Here, we determine whether the M star's radius is  ``inflated'' (i.e. larger than the model-predicted radius) for its inferred mass. Low-mass stellar evolutionary models predict radii and effective temperatures that are $5-15\%$ smaller and hotter, respectively, than observed values (see, e.g., \citealt{Torres:2010,Kraus2011,Birkby2012,Mann2015}). These discrepancies appear to be larger for M dwarfs with fully convective interiors and smaller for more massive M dwarfs with partially convective interiors (\citealt{Han2017}), but this trend does not hold for every system (e.g. \citealt{Han2019}). These effects are seen even in young, pre-main sequence M stars, such as USco 5 \citep{Kraus2015}. Moreover, robust determination of these discrepancies is hindered in part by inaccurate or imprecise measurements of the stellar parameters (see, e.g., \citealt{Healy2019}, whose updated masses for NSVS 07394765, from high-resolution spectroscopy, reduce the ``hyperinflation'' seen by \citet{Cakirli2013} or derived from heterogeneous analysis and observational methods (see \citealt{Torres2013} for a discussion on these issues). 

\begin{figure*}
    \centering
    \includegraphics[width=1.\linewidth]{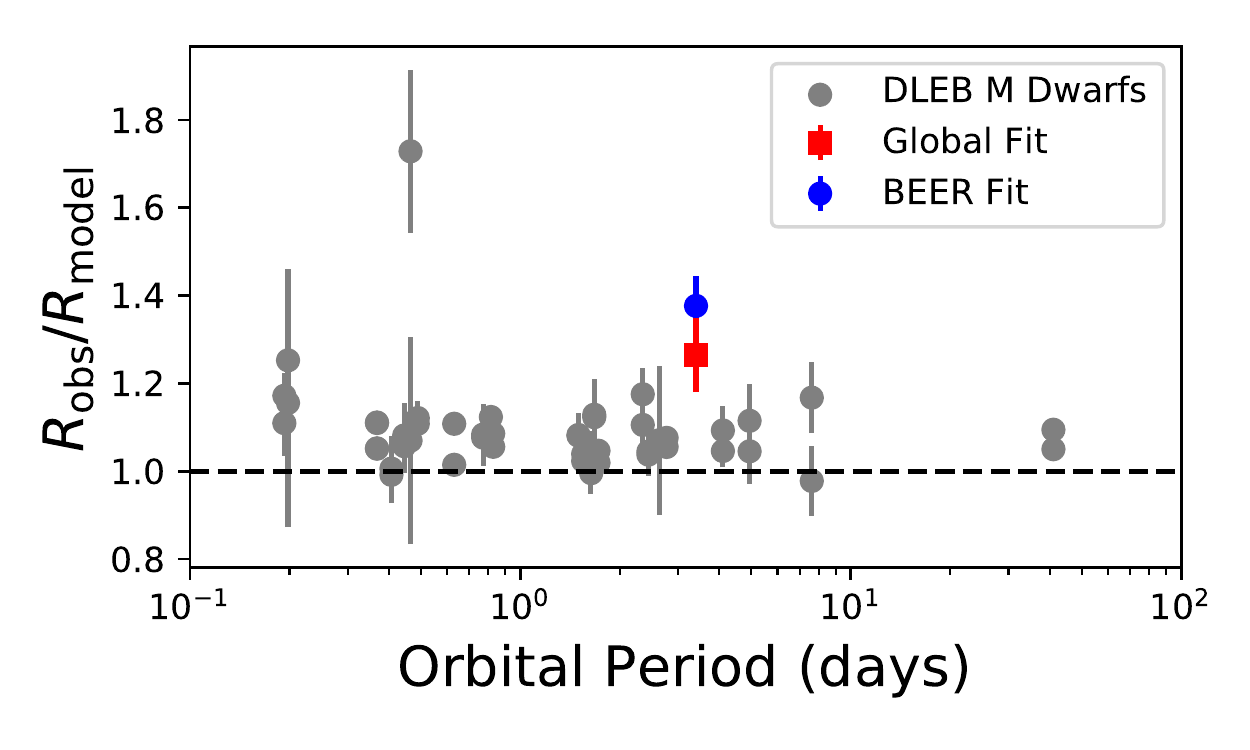}
    \caption{Radius inflation for \thisstar B as inferred from the global fit (red square) and BEER fit (blue point), compared to literature values for M dwarfs in double-lined EBs (grey points) as a function of orbital period. The constraint on the radius inflation from the global fit is comparable to all but the most precise determinations from M-M EBs.}
    \label{fig:rinf2}
\end{figure*}

In Figure \ref{fig:rinf2}, we plot the ratio of observed radii to model-predicted radii for M-M double-lined eclipsing binaries (DLEBs) in the literature, taken from the Detached Eclipsing Binary Catalogue (DEBCat; \citealt{Southworth2015}). We plot the same for \thisstar B, using the radii from both the global and the BEER analysis and comparing to masses and radii from a 200 Myr \citet{Baraffe:2015} isochrone, which is close to the age upper limit inferred for \thisstar A from the global fit. We find that the global-fit radius is $26\pm 8\%$ larger than the 200 Myr isochrone would predict, indicating significant observed-radius inflation relative to the model-predicted value. This larger-than-expected radius is not likely to be a systematic effect caused by our treatment of gravity darkening in the TESS light curve. \citet{Barnes2009} shows that, while a mistreatment of gravity darkening can bias the inferred companion radius if the impact parameter and/or stellar obliquity are large, the effect is smaller for a more aligned transit.

Since we determine quite precise quantities from the light curve -- 1\% fractional uncertainty on the radius ratio, for example -- and since our inferred value of $\beta$ is smaller than one might expect for a hot, rapidly rotating star, we examine how different values of $\beta$ may affect our radius ratio measurement. We calculate the marginalized posterior distribution of $R_2/R_1$ for $\beta = 0.15 \pm 0.02$ from the MCMC chains and determine the median value and 68\% confidence interval. We find that $R_2/R_1 = 0.09765_{-0.00076}^{+0.00074}$; both the value itself and the fractional precision are in strong agreement with the result from the global fit, $R_2/R_1 = 0.0977 \pm 0.008$.

Similarly \thisstar B's radius inferred from the separate BEER analysis is $38\pm 7\%$ larger, driven by the smaller mass but similarly sized radius.
Despite the large mass uncertainty (8\%) on the global analysis value, the precision we achieve on the radius inflation for \thisstar B is comparable to the achieved precision for all but the best-characterized DLEB M dwarfs.

\begin{figure}
    \centering
    \includegraphics[width=1\linewidth]{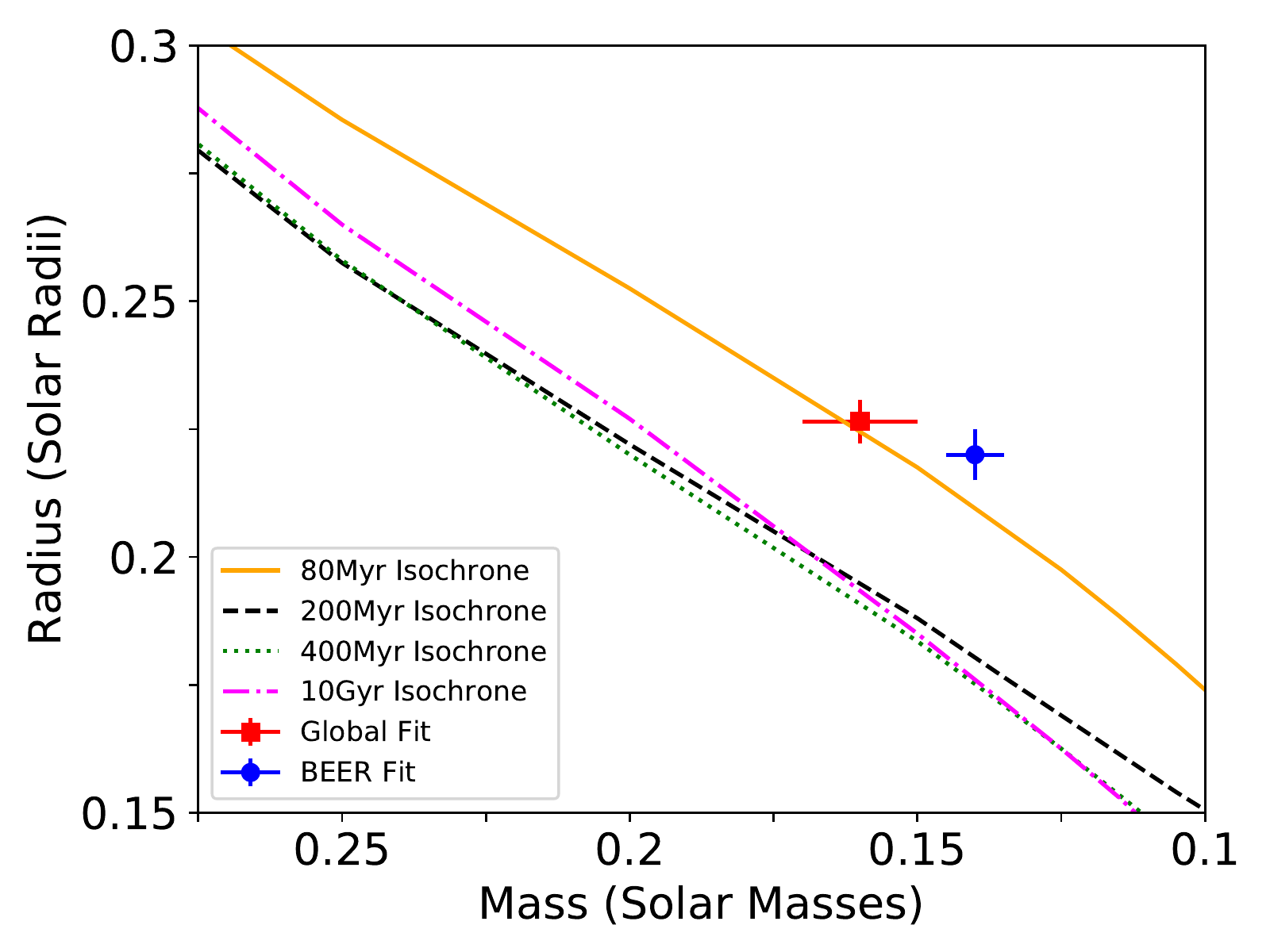}
    \caption{Stellar radius versus mass for \thisstar B as inferred from the global fit (blue circle) and the BEER analysis (red square) and compared to \citet{Baraffe:2015} isochrones over a range of ages. Both analyses' results lie above the 200Myr isochrone (black dotted line), indicative of larger-than-expected radii.}
    \label{fig:rinf}
\end{figure}

To determine if the radius inflation seen in Figure \ref{fig:rinf2} could be an effect of stellar evolution, we compare \thisstar B's mass and radius from the global fit and BEER analyses to isochrones spanning 80 Myr to 10 Gyr from \citet{Baraffe:2015}. As shown in Figure \ref{fig:rinf}, \thisstar B sits above the $\geq200$ Myr isochrones, and the global-fit derived parameters are consistent with the values from the 80 Myr isochrone. This would suggest that either the age inferred from the global fit is $>3\sigma$ too old, or the age is accurate but irradiation effects are inflating \thisstar B's radius. While \thisstar B also sits above the 10 Gyr isochrone, the age of the Universe is approximately 14 Gyr, so we can rule out \thisstar B being an exceptionally old, evolved star.

To determine whether or not the observed radius inflation could be explained by stellar activity (e.g. magnetic inhibition), we estimate the expected radius discrepancy assuming that \thisstar B is tidally synchronized. We calculate the rotation period $P_{\rm rot} = P_{\rm orb} \times (1-e)^{1.5}$, using the median values from Section \ref{sec:GlobalFit}. From $P_{\rm rot}$ and Equation 3 of \citet{West:2015}, we calculate a fractional $H_{\alpha}$ luminosity. We then determine the fractional X-ray luminosity from the fractional $H_{\alpha}$ luminosity via the empirical relation of \citet{Stassun:2012}. Finally, we use the \citet{Stassun:2012} relations between fractional X-ray luminosity and observed-versus-modeled radius discrepancy. From this, we would expect \thisstar B's radius to be inflated by $13\pm 11\%$, which is consistent with our global fit result at $<1.5 \sigma$.

Given \thisstar B's high surface gravity, we do not suspect that \thisstar A's irradiation is inflating \thisstar B's radius. Simplified models of irradiated low-mass stars indicate that the insolation received by \thisstar B would inflate its radius by at most 5\% \citep{Lucy2017}. Additionally, the \thisstar B's ``dayside'' brightness temperature, which is calculated from the reflection-dominated secondary eclipse depth, is approximately its equilibrium temperature ($\sim 4300$K). Thus, most of the incident flux is being reflected or locally reprocessed.

The radius discrepancy exhibited by \thisstar B is similar to that seen in some other young low-mass stars, however. For example, the measured radius of the pre-main sequence ($\sim$800 Myr-old) M-type star PTFEB132.707+19.810B is 20\% larger than predicted by isochrones \citep{Kraus2017}. \citet{Jackson2009} found that isochrones under-predicted the radii of stars in NGC 2516 (age $\sim 150$ Myr) by up to 50\% at an inferred mass of $\sim 0.2 \msun$.

It is possible that the radius discrepancy is the result of an accretion history that is not well-represented by these evolutionary models. \citet{Baraffe2017} showed that episodic accretion models can produce a spread in luminosity at a given temperature and age, which would correspond to a spread in radius. These models attempt to explain the spread seen in FU Orionis stars and are limited to ages below 50 Myr, so it is unclear what effect \thisstar B's accretion history -- and how it may deviate from the accretion history of a similar, single star due to the presence of \thisstar B -- has on the inferred stellar parameters.

We also note that we are, in effect, comparing the predictions of stellar evolutionary models (specifically, the Geneva isochrones as applied to \thisstar A, which indirectly constrain \thisstar B's parameters) to predictions of other stellar evolutionary models (the Baraffe isochrones). 
As the Geneva models are tailored to hot, massive, rapidly rotating stars, while the Baraffe models are intended for cool, low-mass stars, we can get a sense of the relative accuracy of these models. Without a model-independent determination of \thisstar's physical properties, however, we cannot evaluate the absolute accuracy of these models. 

\begin{figure}
    \centering
    \includegraphics[width=1\linewidth]{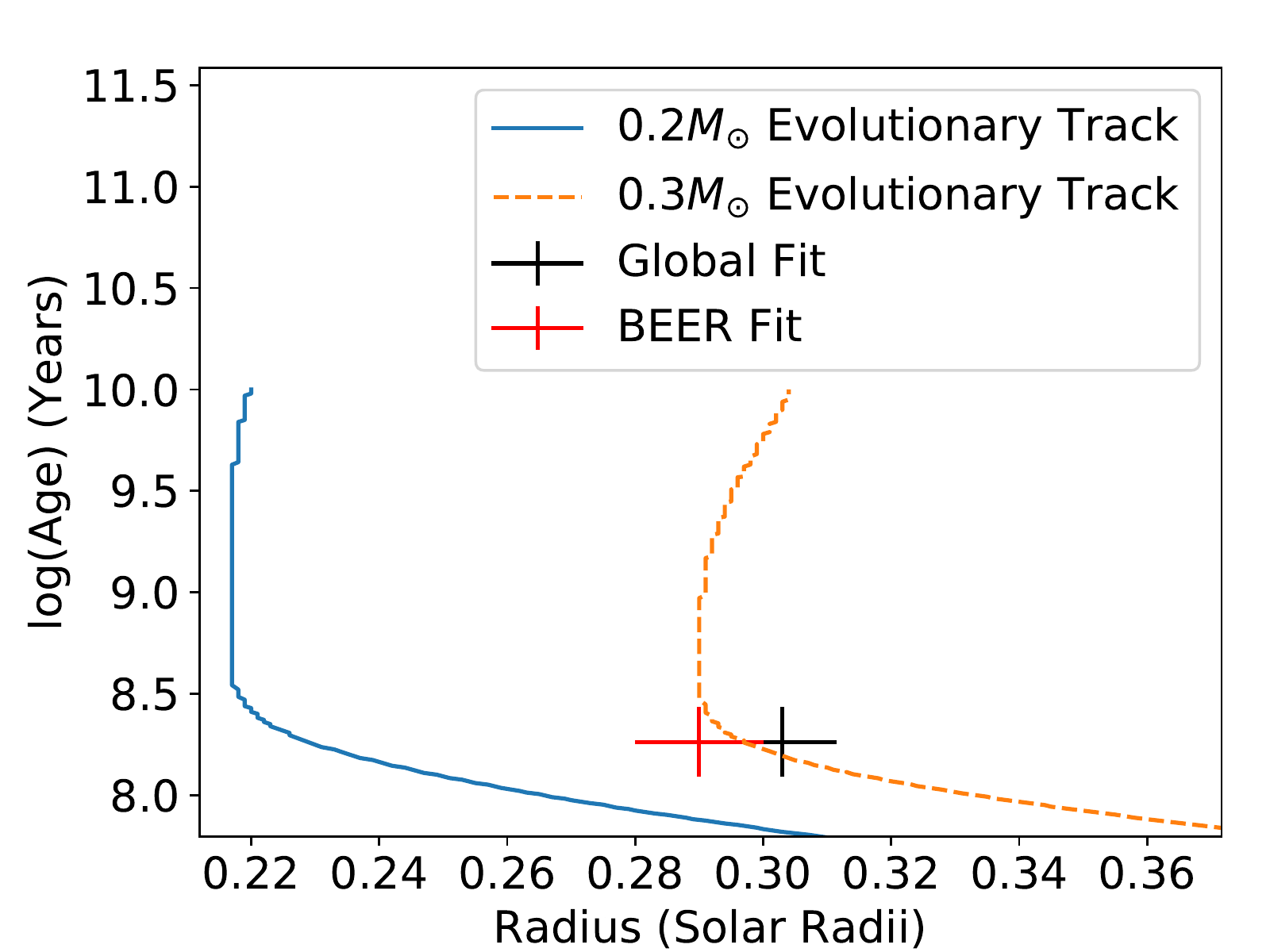}
    \caption{Age versus stellar radius for \thisstar B, using the radius inferred from the global fit (black cross) and the separate BEER analysis (red cross). The lines denote \citet{Baraffe:2015} evolutionary tracks for a $0.2\MS$ (blue line) and $0.3\MS$ (orange line) star. For each track, the main sequence is shown as the nearly-vertical portion of the track.}
    \label{fig:mevol}
\end{figure}

Finally, Figure \ref{fig:mevol} shows that our inferred age for \thisstar\ B is consistent with it being a pre-main sequence star, per \citet{Baraffe:2015} evolutionary models for a 0.2\MS star and a 0.3\MS star. From this evolutionary track, \thisstar B would reach the main sequence at an approximate age of 300 Myr, which is about when \thisstar A would cease burning hydrogen in its core.

\section{Discussion}\label{sec:discussion}

\subsection{Low-mass Companions to Intermediate-mass Stars}
We report the discovery and analysis of the very low mass ratio, $q \equiv M_2/M_1 = 0.07$, binary \thisstar.  Such close-in, low-mass companions to intermediate-mass stars are rare. \citet{DeRosa:2014} found that, for binaries with an A-star primary and relatively small projected separations (between 30 and 125 au), the mass ratio distribution is nearly flat; however, only 18 binaries fell within this projected separation range, and the sample was limited to $q \gtrsim 0.15$. \citet{Moe:2017} examined binaries with $q \geq 0.1$ and found that binaries with B-type primaries and periods less than 20 days prefer a mass ratio $q \approx 0.5$, with low eccentricities, $e \lesssim 0.4$; similarly, \citet{Gullikson:2016} find a mass ratio distribution that peaks at $q \approx 0.3$ for binaries with A- or B-type primaries. However, their sample does not probe separations as short or mass ratios as small as those of \thisstar. As such, it is difficult to measure the shape of the low mass-ratio end of the distribution. The discovery of \thisstar~B demonstrates that the combination of ground-based transit surveys and TESS are capable of finding short-period, eclipsing EMRBs with intermediate-mass primaries. Although beyond the scope of the present work, a comprehensive survey using these facilities could provide the first solid constraints on the mass ratio distribution for such binaries.

Systems such as \thisstar\ can provide unique insights into binary star formation processes. Most binaries are thought to form via core fragmentation (e.g. \citealt{Boss:1979,Boss:1986,Bate:1995}), but some might form through disk fragmentation (\citealt{Kratter:2006,Stamatellos:2011,Mercer:2017}). Such fragmentation is expected to occur at larger separations of 50-200 au from the primary. In either scenario, \thisstar B would likely have migrated in to its present-day orbit; if it migrated during \thisstar A's  pre-main sequence phase, we might expect \thisstar B to have accreted disk material along the way and have a mass that is more similar to \thisstar A. Dynamical interactions with a tertiary star (such as the Lidov-Kozai mechanism; \citealt{Lidov:1962,Kozai1962}) could also produce a short present-day binary separation \citep{Moe:2017}, but we see no evidence for a tertiary in the system. Alternatively, it may be possible for a gaseous clump to migrate to within 10 au of the primary protostar and form a low-mass stellar companion \citep{Meyer:2018}.

Finally, we note that the radius, stellar inclination, and projected rotational velocity of \thisstar A that we infer from our global analysis implies a rotation period of approximately 0.8 days. Some studies of F-M binaries (e.g. \citealt{Fernandez:2009,Chaturvedi:2018}) assume tidal synchronization of the primary star to obtain the M dwarf's mass and radius in a ``model-independent'' way. As the extreme example of \thisstar\ shows and as \citet{Fernandez:2009} found for one of their F-M EBs, this assumption may not hold for stars that experience inefficient tidal braking due to the absence of a substantial convective envelope, and caution must be exercised when applying this assumption to main-sequence stars above the Kraft break at $\teff \approx 6250$ K.

\subsection{Potential Sources of Phase-Curve Analysis Inaccuracies}\label{subsec:beerprobs} 

\subsubsection{Ellipsoidal Variations:} The ellipsoidal variation description assumes a strong equilibrium tide, and such tides are weak in hot stars with radiative envelopes \citep{Zahn:1977}. Moreover, following \citet{Hut:1980}, tidal equilibrium requires a coplanar, circular orbit with corotating stars. \thisstar\ satisfies the first two conditions: however, based on our best-fit values of $R_1, v\sin I_{*,1}$, and $I_{*,1}$, we find a rotation period of $P_{\rm rot,1} \approx 0.9$ days, or $\sim 25\%$ of the orbital period. Although our analytic equilibrium-tide model provides a good fit to the TESS phase curve, we cannot, strictly speaking, assume that \thisstar\ is in tidal equilibrium. 

Dynamically excited tidal oscillations are seen in other eclipsing binaries containing hot stars. Of note are the so-named ``heartbeat stars'' (e.g. \citealt{Kumar:1995,Fuller:2012}) in which a binary companion on an eccentric orbit induces a strong, echocardiogram-like flux variation during periastron passage. This close approach excites a pulsation mode in the primary star whose period is approximately the pseudo-synchronization period (or a harmonic thereof).

Additionally, while we do incorporate gravity-darkening parameters in our phase curve analyses, our calculations in this section do not account for the complex geometry of the primary star's surface. Ignoring any effects due to spin-orbit misalignment, then our choice of which primary stellar radius we use should affect what companion radius we infer: from Section \ref{sec:GlobalFit}, \thisstar A has a 6\% pole-to-equator radius difference. From Equation 2, $M_2 \propto R_1^{-3}$, holding all other quantities fixed. Thus, $\sigma_{M_2}/{M_2} \propto 3\sigma_{R_1}/{R_1}$, and a 6\% change in the input B-star radius can effect an 18\% change in the inferred M star mass.

\subsubsection{Doppler Beaming:} Our calculation of the beaming parameter $\alpha_{\rm beam}$ relies on stellar atmosphere models -- more specifically, on the slope of the stellar SEDs within and near the TESS bandpass. While we use rotating stellar models to calculate disk-integrated fluxes in our global analysis, the NextGen model atmospheres \citep{Hauschildt:1999} we employ in this analysis are agnostic towards rotation and spin-axis-projection effects on the inferred stellar flux. In our global analysis, we find that \thisstar A's rotation axis is roughly perpendicular to our line of sight, so we see more of the gravity-darkened equator than of the brightened poles. In this case, we would over-estimate the surface-integrated flux of the star, likely overestimate $\alpha_{\rm beam}$, and thus likely underestimate the M star's mass. The M star mass inferred from our BEER analysis is indeed smaller than the global fit value, but not significantly so. 

\subsubsection{Reflection/Companion Phase Function:} Ascribing a $\cos{\phi}$ flux variation solely to reflection of incident flux is implausible. While bolometric geometric albedos for stars could be large (between 0.1-1; e.g. \citealt{Rucinski1969,Rucinski:1989}), it is not clear that the geometric albedo in a red-optical filter (such as {\it TESS}) should also be significant. A fully-convective star's atmosphere should not have sufficiently many free electrons to reflect a substantial portion of incident starlight. 

It is more likely that such a phase-variant signal could be due to emitted flux differences across the face of the irradiated star (i.e. the part of the atmosphere being irradiated would be hotter and brighter than the opposing side). Such day-night temperature differences have been inferred from other systems. \citet{Moe2015} provides several examples of EMRBs in the LMC that show a brightening around secondary eclipse, when the dayside of the low-mass companion swings into our line of sight. In addition, the HW Virginis class of variables serves as an extreme example (e.g. \citealt{Almeida2012}.  An HW Vir variable is an eclipsing binary composed of a hot subdwarf star and a low-mass stellar companion in a tight ($P \lesssim 0.5$-day) orbit. In the specific case of the HW Vir variable Konkoly J064029.1+385652.2, \citet{Derekas2015} estimate the temperature at the companion's substellar point to be 22,000K, whereas its effective temperature is a comparatively mild $\sim 4700$K. Irradiation effects have also been studied in the context of mass-transfer binaries \citep{Ritter2000, Hernandez2016}. 

It may also be possible that other scattering processes -- such as Rayleigh and Raman back-scattering off atmospheric molecules -- induce a detectable, phase-dependent flux variation. Assuming incident photons with wavelength at the center of the TESS bandpass ($\lambda = 786.5$nm), and assuming they are incident upon an atmospheric shell of pure VO and/or TiO (with molecular polarizability $\alpha \approx 10$\AA$^3$)\footnote{Computational Chemistry Comparison and Benchmark Database, NIST Standard Reference Database Number 101, Release 20, August 2019, Ed: Russell D. Johnson III, http://cccbdb.nist.gov/} with thickness equal to one scale height, a back-scattering signal of  few-hundred-ppm amplitude may be produced at secondary eclipse (ignoring, of course, the occultation of the star doing the scattering). These polarizability quantities are not empirically determined, and a full numerical analysis of these scattering effects is beyond the scope of this paper.

\section{Conclusions}\label{sec:conclusion}
We report the discovery of an eclipsing binary, \thisstar, with an extreme mass ratio of $q = 0.07$. We characterized \thisstar\ by jointly analyzing the TESS phase curve, ground-based photometric and spectroscopic transits, RVs, the SED, and the Geneva isochrones. Our inferred parameters are consistent with a young EB consisting of a late-B dwarf primary star and a pre-main sequence M star companion. For its inferred mass, the M star's radius is significantly larger than the value predicted by a pre-main sequence isochrone of similar age by $26\pm 8$\%.

We also performed a separate analysis of out-of-eclipse TESS data, fitting for Doppler beaming \citep{Loeb:2003}, ellipsoidal variation, and reflected/re-radiation (BEER; \citealt{Faigler:2011}) effects. We find no significant differences between this analysis and the global fit analysis. In Section \ref{subsec:beerprobs}, however, we offer reasons for why one should not generalize this conclusion to all EBs.

Our discovery of \thisstar\ emphasizes the sensitivity of KELT and other hot-star transit surveys to close-in, cool stellar companions orbiting intermediate-mass stars. These transit surveys can thus improve the census of companions to intermediate-mass stars -- both by pushing to lower mass and flux ratios and shorter orbital separations and by observing eclipses that, along with RVs and other complementary datasets, enable comprehensive characterizations of these systems. Binaries such as \thisstar -- an extreme example of an EMRB in which both stars are themselves at extreme ends of parameter space -- provide strong tests our understanding of EMRB formation, evolution, and characterization.

\section*{Acknowledgments}
The authors thank Avi Shporer and Chelsea Huang for their expertise on analyzing and processing TESS photometry, respectively. The authors also thank the anonymous reviewer for the helpful comments and suggestions. D.J.S. is supported as an Eberly Research Fellow by the Eberly College of Science at the Pennsylvania State University. The Center for Exoplanets and Habitable Worlds is supported by the Pennsylvania State University, the Eberly College of Science, and the Pennsylvania Space Grant Consortium. 

The authors wish to recognize and acknowledge the very significant cultural role and reverence that the summit of Mauna Kea has always had within the indigenous Hawaiian community.  We are most fortunate to have the opportunity to conduct observations from this mountain.

\subsection*{Data Availability}
This project makes use of data from the KELT survey, including support from The Ohio State University, Vanderbilt University, and Lehigh University, along with the KELT follow-up collaboration.

This paper includes data taken at The McDonald Observatory of The University of Texas at Austin. This paper also includes data gathered with the 6.5 meter Magellan Telescopes located at Las Campanas Observatory, Chile.

Some of the data presented herein were obtained at the W.M. Keck Observatory, which is operated as a scientific partnership among the California Institute of Technology, the University of California and the National Aeronautics and Space Administration. These data are available through the Keck Observatory Archive (\url{https://koa.ipac.caltech.edu}). The Observatory was made possible by the generous financial support of the W.M. Keck Foundation.

This publication makes use of data products from the Two Micron All Sky Survey, which is a joint project of the University of Massachusetts and the Infrared Processing and Analysis Center/California Institute of Technology, funded by the National Aeronautics and Space Administration and the National Science Foundation.

This work has made use of data from the European Space Agency (ESA) mission
{\it Gaia} (\url{https://www.cosmos.esa.int/gaia}), processed by the {\it Gaia}
Data Processing and Analysis Consortium (DPAC,
\url{https://www.cosmos.esa.int/web/gaia/dpac/consortium}). Funding for the DPAC
has been provided by national institutions, in particular the institutions
participating in the {\it Gaia} Multilateral Agreement.

This research has made use of NASA's Astrophysics Data System; Astropy,\footnote{http://www.astropy.org} a community-developed core Python package for Astronomy \citep{astropy:2013, astropy:2018}; and the SIMBAD database, operated at CDS, Strasbourg, France \citep{Wenger2000}. 
 
\bibliographystyle{mnras}
\bibliography{KJ06C004063.bib}
\bsp	
\label{lastpage}
\end{document}